\documentclass[10pt,nolinenumbers, twocolumn, superscriptaddress, notitlepage]{revtex4-2}
\usepackage{amsbsy}
\usepackage{amsmath,amssymb}
\usepackage[usenames]{color}
\usepackage[colorlinks,linkcolor=black,citecolor=black]{hyperref}
\usepackage{graphicx}
\usepackage[caption=false]{subfig}
\usepackage{tikz}

\tikzset{
  meas/.style={
    draw,
    rectangle,
    minimum width=6mm,
    minimum height=5mm,
    font=\small,
    inner sep=1pt
  }
}
\newcommand{\cqt}{Centre for Quantum Technologies, National University of Singapore, Singapore 117543}
\newcommand{\uoh}{School of Physics, University of Hyderabad, India 500046}

\begin{document}

\title{Quantum Contextuality and Entanglement-Free Grover Search in a Trapped-Ion Optical Qudit}
\author{Tarun Dutta}
\email{tarunduttaz@gmail.com}
\address{\cqt}
\address{\uoh}
\author{Jasper Phua Sing Cheng}
\address{\cqt}
\author{Alex Jin}
\address{\cqt}

\author{Sergi Ramos-Calderer}

\author{José Ignacio Latorre}
\address{\cqt}

\author{Manas Mukherjee}
\email{manas.mukh@gmail.com}
\address{\cqt}
\begin{abstract}

Quantum computational advantage is generally attributed to coherent interference and other non-classical resources, yet their respective roles remain difficult to disentangle in experimental platforms where multipartite entanglement is inherently present. High-dimensional quantum systems provide an attractive route for investigating these resources while simultaneously reducing hardware overhead for quantum information processing. Here we realize a programmable four-dimensional optical qudit encoded in a single trapped $^{138}\mathrm{Ba}^{+}$ ion and demonstrate universal coherent control through phase-programmable optical rotations. Using this platform, we implement an entanglement-free realization of Grover's quantum search algorithm, achieving target-state identification probabilities of up to $94.5\pm2.0\%$. Within the same processor, we further demonstrate state-dependent quantum contextuality through a Clauser--Horne--Shimony--Holt (CHSH)-type noncontextuality inequality, obtaining a maximum violation of $S = 2.816 \pm 0.082$, in close agreement with the Tsirelson bound. By integrating programmable quantum computation and contextuality measurements within a single multilevel trapped-ion platform, our work establishes a versatile architecture for investigating the relationship between coherent interference and contextuality in quantum information processing and provides a scalable route toward high-dimensional quantum technologies.

\end{abstract}
\maketitle
\section{Introduction}
Understanding the physical resources responsible for quantum computational advantage remains one of the central challenges in quantum information science. Although quantum algorithms exploit coherent superposition and interference, identifying the non-classical features that distinguish quantum computation from efficient classical simulation remains an active area of research. The Gottesman--Knill theorem establishes that quantum circuits composed solely of Clifford operations are efficiently classically simulable, implying that universal quantum computation requires resources beyond stabilizer operations~\cite{Gottesman1998,Aaronson2004}. In particular, quantum contextuality has emerged as an essential resource for magic-state quantum computation and fault-tolerant quantum information processing~\cite{Howard2014}. Despite these theoretical advances, quantum algorithms and contextuality are typically investigated independently, often using different experimental platforms and control architectures. Establishing a common experimental framework capable of probing both quantum computation and its underlying non-classical resources therefore remains an important challenge.

At the same time, scaling quantum information processors presents significant practical challenges. As the number of individually addressable qubits increases, maintaining high-fidelity control becomes increasingly demanding because of calibration overhead, control complexity, and crosstalk arising from spatial or spectral crowding in trapped ions, neutral atoms, and superconducting circuits~\cite{Monroe2021,Preskill2018}. Furthermore, many quantum algorithms rely heavily on entangling gates, which remain among the most resource-intensive operations in existing quantum processors~\cite{NielsenChuang,CiracZoller1995,Monroe2021}. High-dimensional quantum systems (qudits with $d>2$) provide a promising route toward hardware-efficient quantum information processing by encoding a larger computational Hilbert space within a single physical system while enabling native multilevel operations~\cite{Wang2020,Hrmo2023,ringbauer2022,Nikolaeva2024,Zalivako2025Multiqudit,Shi2026,Lanyon2009}. However, scalable coherent control of multilevel systems remains experimentally demanding because arbitrary $SU(d)$ transformations generally require sequences of $O(d^2)$ pairwise transitions~\cite{Muthukrishnan2000}. Among the available platforms, trapped atomic ions are particularly attractive because their long-lived electronic manifolds naturally provide well-isolated multilevel states with excellent coherence properties and high-fidelity optical control~\cite{PhysRevLett.95.060502,Kotler2011,PhysRevLett.113.220501,Wang2021}.

Beyond their hardware advantages, multilevel quantum systems provide a unique setting in which quantum algorithms and foundational tests of quantum mechanics can be implemented within the same physical particle. Lloyd and Meyer showed that Grover's quantum search can be realized through coherent interference in a single multilevel quantum system rather than entanglement between physically distinct qubits~\cite{Lloyd1999,Meyer2000}. Likewise, quantum contextuality can be investigated in high-dimensional Hilbert spaces using state-dependent noncontextuality inequalities. Four-dimensional Hilbert spaces are especially attractive because they simultaneously constitute genuine qudits while admitting an effective two-qubit encoding. This dual description enables programmable multilevel quantum computation and Bell-type contextuality tests to be implemented using a common physical system and control toolbox. Although these capabilities have been explored separately, an experimental platform combining universal qudit control, quantum algorithm implementation, and contextuality measurements within the same programmable architecture has remained lacking. More importantly, such a platform enables quantum computational performance and experimentally measured non-classicality to be benchmarked within the same physical system, providing an opportunity to investigate whether these quantities are directly connected.

Here we realize a programmable four-dimensional optical qudit encoded in the metastable-state manifold of a single trapped $^{138}\mathrm{Ba}^{+}$ ion. Universal coherent control of the four-dimensional Hilbert space is achieved through phase-programmable optical rotations acting on embedded $SU(2)$ subspaces, enabling arbitrary $SU(4)$ operations. Using this programmable processor, we implement Grover's quantum search entirely within the Hilbert space of a single qudit through coherent multilevel interference, achieving target-state identification probabilities of up to $94.5\pm2.0\%$. Using the same experimental platform and control framework, we further perform a CHSH-inspired state-dependent noncontextuality test and observe a maximum violation of $S=2.816\pm0.082$, approaching the Tsirelson bound. Together, these results establish a unified trapped-ion platform for programmable multilevel quantum information processing and contextuality studies, providing a route toward experimentally investigating the relationship between non-classical resources and quantum computational performance.
\begin{figure*}[t]
  \centering
  \includegraphics[width=\linewidth]{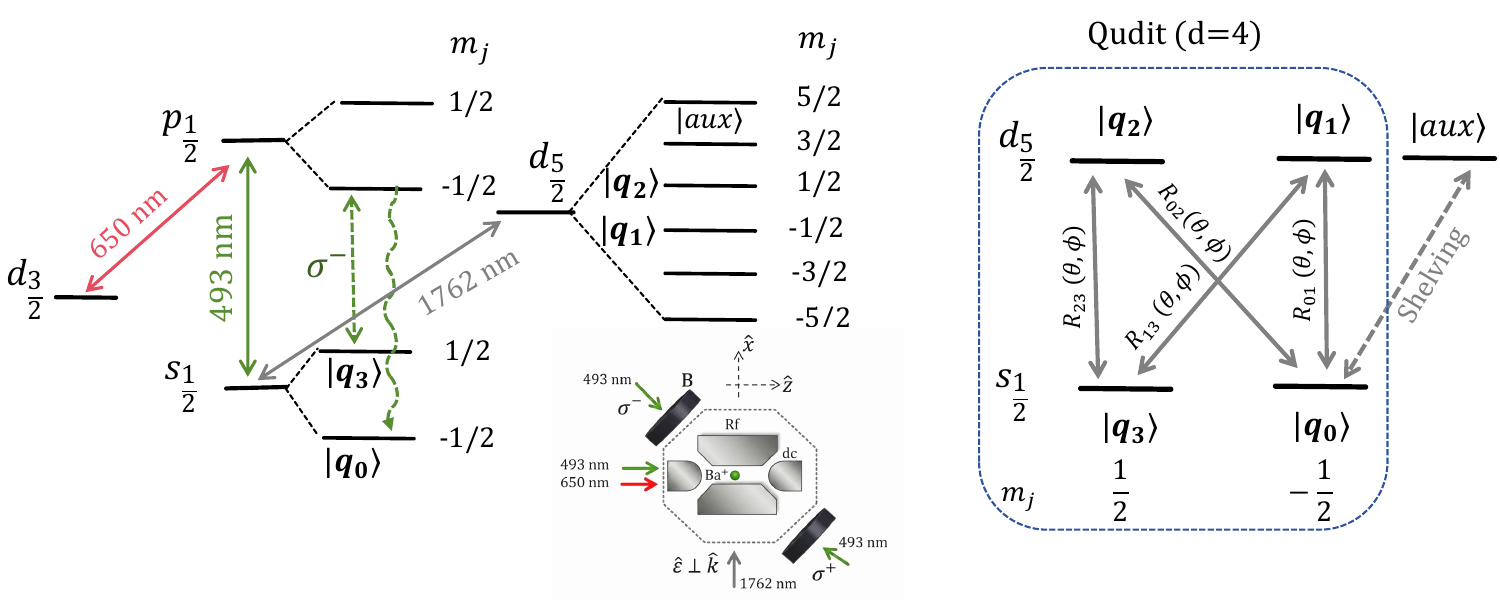}
\caption{
 Energy-level structure of the trapped-ion $^{138}\mathrm{Ba}^{+}$ qudit and the coherent transitions employed for initialization and quantum control. The \(493~\mathrm{nm}\) transition is used for Doppler cooling, while the \(650~\mathrm{nm}\) transition repumps population from the metastable \(d_{3/2}\) manifold back to the cooling cycle. A resonant \(493~\mathrm{nm}\,\sigma^{-}\)-polarized optical-pumping beam drives the \(|s_{1/2},m_J=+1/2\rangle \rightarrow |p_{1/2},m_J=-1/2\rangle\) transition, thereby initializing the ion into the dark state \(|s_{1/2},m_J=-1/2\rangle\equiv|q_0\rangle\). The four-dimensional optical qudit computational basis is encoded in the Zeeman sublevels of the \(s_{1/2}\) and \(d_{5/2}\) electronic manifolds. The basis states \(\{|q_0\rangle, |q_1\rangle, |q_2\rangle, |q_3\rangle\}\) are defined using the \(m_J=\pm1/2\) sublevels of the \(s_{1/2}\) ground state and the \(m_J=\pm1/2\) sublevels of the metastable \(d_{5/2}\) state. Coherent single-qudit control is realized using \(1762~\mathrm{nm}\) through the resonant optical rotations \(R_{01}(\theta,\phi)\), \(R_{02}(\theta,\phi)\), \(R_{13}(\theta,\phi)\), and \(R_{23}(\theta,\phi)\), while auxiliary Zeeman states provide additional pathways for shelving, and high-fidelity quantum manipulation. (center-bottom) Schematic of the trapped-ion experimental setup showing the orientation of the trap and the laser beam geometry used to cool, pump, and prepare the $^{138}\mathrm{Ba}^{+}$ ion. A single $^{138}\mathrm{Ba}^{+}$ ion is confined at the trap center. A pair of permanent magnets in a Helmholtz configuration generates a static magnetic field ($B$), defining the quantization axis.
}
\label{qudit}
\end{figure*}
\vspace{-5mm}
\section{Results}
\subsection{Programmable trapped-ion optical qudit}
We realize a programmable four-dimensional optical qudit using a single
$^{138}\mathrm{Ba}^{+}$ ion confined in a blade type Paul trap
(Fig.~\ref{qudit})~\cite{Dutta2020NoiseProbe,DuttaClassifier, Dutta2025HybridTraining}. The computational basis is encoded in two Zeeman sublevels of the
$6S_{1/2}$ ground-state manifold and two Zeeman sublevels of the metastable
$5D_{5/2}$ manifold, 
$\mathcal{B}=
\left\{
|q_0\rangle,
|q_1\rangle,
|q_2\rangle,
|q_3\rangle
\right\},$
%
%
forming a four-dimensional Hilbert space for programmable quantum information processing, illustrated in the Fig.~\ref{qudit}. State initialization is achieved by optical pumping on the $493~\mathrm{nm}$ transition, while coherent manipulation is performed on the narrow-linewidth
$1762~\mathrm{nm}$ electric-quadrupole transition. Projective measurement is realized through state-selective shelving followed by fluorescence detection, with an auxiliary metastable level used exclusively for readout and therefore excluded from the computational basis.

Although the selected Zeeman manifold supports ten optical couplings, only four are required for coherent control,
\begin{equation}
|q_0\rangle\leftrightarrow|q_1\rangle,\quad
|q_0\rangle\leftrightarrow|q_2\rangle,\quad
|q_1\rangle\leftrightarrow|q_3\rangle,\quad
|q_2\rangle\leftrightarrow|q_3\rangle,
\label{eq:transition_set}
\end{equation}
which were chosen to maximize optical coupling strength while minimizing magnetic-field sensitivity and off-resonant excitation. These experimentally accessible transitions define the coherent-control toolbox employed throughout this work.

The corresponding connectivity is described by the adjacency matrix

\begin{equation}
M=
\begin{pmatrix}
0&1&1&0\\
1&0&0&1\\
1&0&0&1\\
0&1&1&0
\end{pmatrix},
\label{eq:adjacency}
\end{equation}

whose non-zero off-diagonal elements denote directly addressable optical transitions. Although sparse, the interaction graph is fully connected, enabling arbitrary population transfer between any pair of basis states through finite sequences of experimentally accessible operations.

The four-level manifold is naturally isomorphic to an effective two-qubit computational basis,
\begin{equation}
|q_0\rangle\equiv|00\rangle,\quad
|q_1\rangle\equiv|01\rangle,\quad
|q_2\rangle\equiv|10\rangle,\quad
|q_3\rangle\equiv|11\rangle,
\label{eq:qubit_mapping}
\end{equation}

allowing the same physical processor to implement both native qudit protocols and qubit-equivalent quantum algorithms. Throughout this work, this correspondence provides a unified framework for universal coherent control, an entanglement-free realization of Grover's search algorithm, and CHSH-type contextuality measurements within a single trapped ion.


 A key feature of the present architecture is the symmetric choice of transition pairs with opposite or nearly symmetric Zeeman sensitivities, leading to partial cancellation of magnetic-field-induced frequency shifts and reduced first-order dephasing. The selected geometry also suppresses differential AC Stark shifts and spectator-state coupling by restricting each operation to the relevant transition manifold, unlike architectures that simultaneously couple multiple qudit states through a common reference level~\cite{Nikolaeva2024, Shi2026}. This enables stable, high-fidelity sequential control within the four-dimensional Hilbert space.

\subsubsection*{Universal coherent \texorpdfstring{$SU(4)$}{SU(4)} control}
Universal coherent control of the four-dimensional Hilbert space is achieved by combining resonant optical rotations acting on experimentally accessible two-level subspaces. Each pulse implements an embedded $SU(2)$ operation within the $SU(4)$ manifold,

\begin{equation}
U_{ij}(\theta,\phi)=
\exp\!\left[
-\frac{i\theta}{2}
\left(
\cos\phi\,\lambda_x^{(ij)}
+
\sin\phi\,\lambda_y^{(ij)}
\right)
\right],
\label{eq:embedded_rotation}
\end{equation}

where $\theta=\Omega t$ is the pulse area, $\phi$ is the optical phase, and
$\lambda_{x,y}^{(ij)}$ denote generalized Pauli operators acting only on the
$\{|q_i\rangle,|q_j\rangle\}$ subspace.

Experimentally, coherent control is implemented using the four directly
accessible optical transitions

\[
\mathcal{G}
=
\{
R_{01},
R_{02},
R_{13},
R_{23}
\},
\]

from which arbitrary coherent operations are synthesized as

\begin{equation}
U=
\prod_{k=1}^{N}
R_{\alpha_k}(\theta_k,\phi_k),
\qquad
\alpha_k\in
\{01,02,13,23\},
\label{eq:gate_sequence}
\end{equation}

where each factor denotes one experimentally applied resonant pulse. Logical
operations correspond to particular pulse sequences within this elementary gate
set, while arbitrary multilevel transformations are obtained by combining pulse
phases, durations, and ordering. This universal control framework forms the basis for both the contextuality measurements and the entanglement-free implementation of Grover's search algorithm presented below.
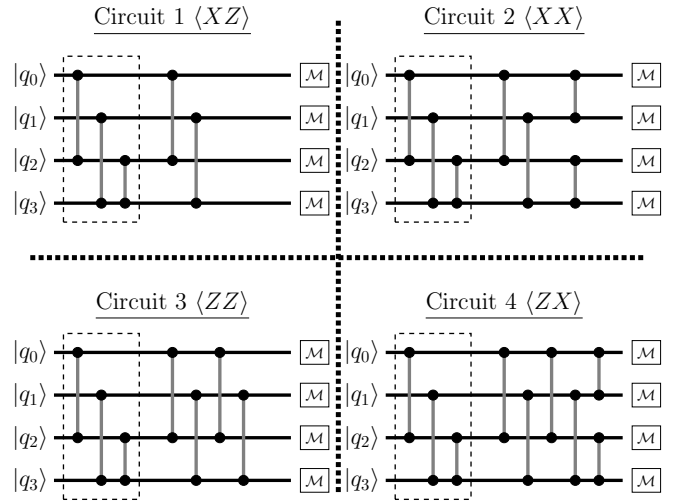
\begin{figure}[htbp]
\centering
\resizebox{\columnwidth}{!}{%
\begin{tikzpicture}
    \node [font=\fontsize{14}{1}\selectfont] at (2.5,5) {\underline{Circuit 1 $\langle XZ \rangle$}} ; 
    \draw[black, line width = 2pt] (0,3.85) -- (5,3.85);
    \draw[black, line width = 2pt] (0,2.95) -- (5,2.95);
    \draw[black, line width = 2pt] (0,2.05) -- (5,2.05);
    \draw[black, line width = 2pt] (0,1.15) -- (5,1.15);
    
    \draw[gray, line width = 2pt] (0.5,3.85) -- (0.5,2.05);
    \draw[gray, line width = 2pt] (1.0,2.95) -- (1.0,1.15);
    \draw[gray, line width = 2pt] (1.5,2.05) -- (1.5,1.15);

    \draw[gray, line width = 2pt] (2.5,3.85) -- (2.5,2.05);
    \draw[gray, line width = 2pt] (3.0,2.95) -- (3.0,1.15);

    \filldraw (0.5,3.85) circle (3pt); \filldraw (0.5,2.05) circle (3pt);
    \filldraw (1.0,2.95) circle (3pt); \filldraw (1.0,1.15) circle (3pt);
    \filldraw (1.5,2.05) circle (3pt); \filldraw (1.5,1.15) circle (3pt);
    \filldraw (2.5,3.85) circle (3pt); \filldraw (2.5,2.05) circle (3pt);
    \filldraw (3.0,2.95) circle (3pt); \filldraw (3.0,1.15) circle (3pt);

    \draw[dashed, black, line width = 0.8pt] (0.2, 0.75) rectangle (1.8, 4.25);

    \node [font=\fontsize{14}{1}\selectfont] at (-0.5,3.85) {$|q_0\rangle$} ; 
    \node [font=\fontsize{14}{1}\selectfont] at (-0.5,2.95) {$|q_1\rangle$} ; 
    \node [font=\fontsize{14}{1}\selectfont] at (-0.5,2.05) {$|q_2\rangle$} ; 
    \node [font=\fontsize{14}{1}\selectfont] at (-0.5,1.15) {$|q_3\rangle$} ; 
    
    \node[meas] at (5.5,3.85) {$\mathcal{M}$};
    \node[meas] at (5.5,2.95) {$\mathcal{M}$};
    \node[meas] at (5.5,2.05) {$\mathcal{M}$};
    \node[meas] at (5.5,1.15) {$\mathcal{M}$};

    \node [font=\fontsize{14}{1}\selectfont] at (9.5,5) {\underline{Circuit 2 $\langle XX \rangle$}} ; 
    \draw[black, line width = 2pt] (7,3.85) -- (12,3.85);
    \draw[black, line width = 2pt] (7,2.95) -- (12,2.95);
    \draw[black, line width = 2pt] (7,2.05) -- (12,2.05);
    \draw[black, line width = 2pt] (7,1.15) -- (12,1.15);

    \draw[gray, line width = 2pt] (7.5,3.85) -- (7.5,2.05);
    \draw[gray, line width = 2pt] (8.0,2.95) -- (8.0,1.15);
    \draw[gray, line width = 2pt] (8.5,2.05) -- (8.5,1.15);

    \draw[gray, line width = 2pt] (9.5,3.85) -- (9.5,2.05);
    \draw[gray, line width = 2pt] (10.0,2.95) -- (10.0,1.15);
    \draw[gray, line width = 2pt] (11.0,3.85) -- (11.0,2.95);
    \draw[gray, line width = 2pt] (11.0,2.05) -- (11.0,1.15);

    \filldraw (7.5,3.85) circle (3pt); \filldraw (7.5,2.05) circle (3pt);
    \filldraw (8.0,2.95) circle (3pt); \filldraw (8.0,1.15) circle (3pt);
    \filldraw (8.5,2.05) circle (3pt); \filldraw (8.5,1.15) circle (3pt);
    \filldraw (9.5,3.85) circle (3pt); \filldraw (9.5,2.05) circle (3pt);
    \filldraw (10.0,2.95) circle (3pt); \filldraw (10.0,1.15) circle (3pt);
    \filldraw (11.0,3.85) circle (3pt); \filldraw (11.0,2.95) circle (3pt);
    \filldraw (11.0,2.05) circle (3pt); \filldraw (11.0,1.15) circle (3pt);

    \draw[dashed, black, line width = 0.8pt] (7.2, 0.75) rectangle (8.8, 4.25);

    \node [font=\fontsize{14}{1}\selectfont] at (6.5,3.85) {$|q_0\rangle$} ; 
    \node [font=\fontsize{14}{1}\selectfont] at (6.5,2.95) {$|q_1\rangle$} ; 
    \node [font=\fontsize{14}{1}\selectfont] at (6.5,2.05) {$|q_2\rangle$} ; 
    \node [font=\fontsize{14}{1}\selectfont] at (6.5,1.15) {$|q_3\rangle$} ; 

    \node[meas] at (12.5,3.85) {$\mathcal{M}$};
    \node[meas] at (12.5,2.95) {$\mathcal{M}$};
    \node[meas] at (12.5,2.05) {$\mathcal{M}$};
    \node[meas] at (12.5,1.15) {$\mathcal{M}$};

    \draw[black, dotted, line width = 3pt] (-0.5,0) -- (12.5,0);
    \draw[black, dotted, line width = 3pt] (6,5) -- (6,-5);

    \node [font=\fontsize{14}{1}\selectfont] at (2.5,-1) {\underline{Circuit 3 $\langle ZZ \rangle$}} ; 
    \draw[black, line width = 2pt] (0,-2) -- (5,-2);
    \draw[black, line width = 2pt] (0,-2.9) -- (5,-2.9);
    \draw[black, line width = 2pt] (0,-3.8) -- (5,-3.8);
    \draw[black, line width = 2pt] (0,-4.7) -- (5,-4.7);

    \draw[gray, line width = 2pt] (0.5,-2) -- (0.5,-3.8);
    \draw[gray, line width = 2pt] (1.0,-2.9) -- (1.0,-4.7);
    \draw[gray, line width = 2pt] (1.5,-3.8) -- (1.5,-4.7);

    \draw[gray, line width = 2pt] (2.5,-2) -- (2.5,-3.8);
    \draw[gray, line width = 2pt] (3.0,-2.9) -- (3.0,-4.7);
    \draw[gray, line width = 2pt] (3.5,-2) -- (3.5,-3.8);
    \draw[gray, line width = 2pt] (4.0,-2.9) -- (4.0,-4.7);

    \filldraw (0.5,-2) circle (3pt);   \filldraw (0.5,-3.8) circle (3pt);
    \filldraw (1.0,-2.9) circle (3pt); \filldraw (1.0,-4.7) circle (3pt);
    \filldraw (1.5,-3.8) circle (3pt); \filldraw (1.5,-4.7) circle (3pt);
    \filldraw (2.5,-2) circle (3pt);   \filldraw (2.5,-3.8) circle (3pt);
    \filldraw (3.0,-2.9) circle (3pt); \filldraw (3.0,-4.7) circle (3pt);
    \filldraw (3.5,-2) circle (3pt);   \filldraw (3.5,-3.8) circle (3pt);
    \filldraw (4.0,-2.9) circle (3pt); \filldraw (4.0,-4.7) circle (3pt);

    \draw[dashed, black, line width = 0.8pt] (0.2, -5.1) rectangle (1.8, -1.6);

    \node [font=\fontsize{14}{1}\selectfont] at (-0.5,-2) {$|q_0\rangle$} ; 
    \node [font=\fontsize{14}{1}\selectfont] at (-0.5,-2.9) {$|q_1\rangle$} ; 
    \node [font=\fontsize{14}{1}\selectfont] at (-0.5,-3.8) {$|q_2\rangle$} ; 
    \node [font=\fontsize{14}{1}\selectfont] at (-0.5,-4.7) {$|q_3\rangle$} ; 

    \node[meas] at (5.5,-2) {$\mathcal{M}$};
    \node[meas] at (5.5,-2.9) {$\mathcal{M}$};
    \node[meas] at (5.5,-3.8) {$\mathcal{M}$};
    \node[meas] at (5.5,-4.7) {$\mathcal{M}$};

    \node [font=\fontsize{14}{1}\selectfont] at (9.5,-1) {\underline{Circuit 4 $\langle ZX \rangle$}} ; 
    \draw[black, line width = 2pt] (7,-2) -- (12,-2);
    \draw[black, line width = 2pt] (7,-2.9) -- (12,-2.9);
    \draw[black, line width = 2pt] (7,-3.8) -- (12,-3.8);
    \draw[black, line width = 2pt] (7,-4.7) -- (12,-4.7);

    \draw[gray, line width = 2pt] (7.5,-2) -- (7.5,-3.8);
    \draw[gray, line width = 2pt] (8.0,-2.9) -- (8.0,-4.7);
    \draw[gray, line width = 2pt] (8.5,-3.8) -- (8.5,-4.7);

    \draw[gray, line width = 2pt] (9.5,-2) -- (9.5,-3.8);
    \draw[gray, line width = 2pt] (10.0,-2.9) -- (10.0,-4.7);
    \draw[gray, line width = 2pt] (10.5,-2) -- (10.5,-3.8);
    \draw[gray, line width = 2pt] (11.0,-2.9) -- (11.0,-4.7);
    \draw[gray, line width = 2pt] (11.5,-2) -- (11.5,-2.9);
    \draw[gray, line width = 2pt] (11.5,-3.8) -- (11.5,-4.7);

    \filldraw (7.5,-2) circle (3pt);   \filldraw (7.5,-3.8) circle (3pt);
    \filldraw (8.0,-2.9) circle (3pt); \filldraw (8.0,-4.7) circle (3pt);
    \filldraw (8.5,-3.8) circle (3pt); \filldraw (8.5,-4.7) circle (3pt);
    \filldraw (9.5,-2) circle (3pt);   \filldraw (9.5,-3.8) circle (3pt);
    \filldraw (10.0,-2.9) circle (3pt); \filldraw (10.0,-4.7) circle (3pt);
    \filldraw (10.5,-2) circle (3pt);  \filldraw (10.5,-3.8) circle (3pt);
    \filldraw (11.0,-2.9) circle (3pt); \filldraw (11.0,-4.7) circle (3pt);
    \filldraw (11.5,-2) circle (3pt);  \filldraw (11.5,-2.9) circle (3pt);
    \filldraw (11.5,-3.8) circle (3pt); \filldraw (11.5,-4.7) circle (3pt);

    \draw[dashed, black, line width = 0.8pt] (7.2, -5.1) rectangle (8.8, -1.6);

    \node [font=\fontsize{14}{1}\selectfont] at (6.5,-2) {$|q_0\rangle$} ; 
    \node [font=\fontsize{14}{1}\selectfont] at (6.5,-2.9) {$|q_1\rangle$} ; 
    \node [font=\fontsize{14}{1}\selectfont] at (6.5,-3.8) {$|q_2\rangle$} ; 
    \node [font=\fontsize{14}{1}\selectfont] at (6.5,-4.7) {$|q_3\rangle$} ; 

    \node[meas] at (12.5,-2) {$\mathcal{M}$};
    \node[meas] at (12.5,-2.9) {$\mathcal{M}$};
    \node[meas] at (12.5,-3.8) {$\mathcal{M}$};
    \node[meas] at (12.5,-4.7) {$\mathcal{M}$};
\end{tikzpicture}%
}
\caption{
  Pulse sequences implementing the four effective CHSH measurement configurations in the single-qudit system. Connected levels represent coherent rotations within the corresponding two-level subspaces. The first three transitions prepare the superposition state \((|q_0\rangle+|q_3\rangle)/\sqrt{2}\), corresponding to the Bell state \((|00\rangle+|11\rangle)/\sqrt{2}\) under the qudit--qubit correspondence. The subsequent transitions implement basis rotations parameterized by the phase angle \(\theta\). Final projection pulses realize the four effective measurement settings used to evaluate the CHSH parameter.
}
\label{CHSH1}
\end{figure}
\begin{figure}[t]
  \centering
  \includegraphics[width=\linewidth]{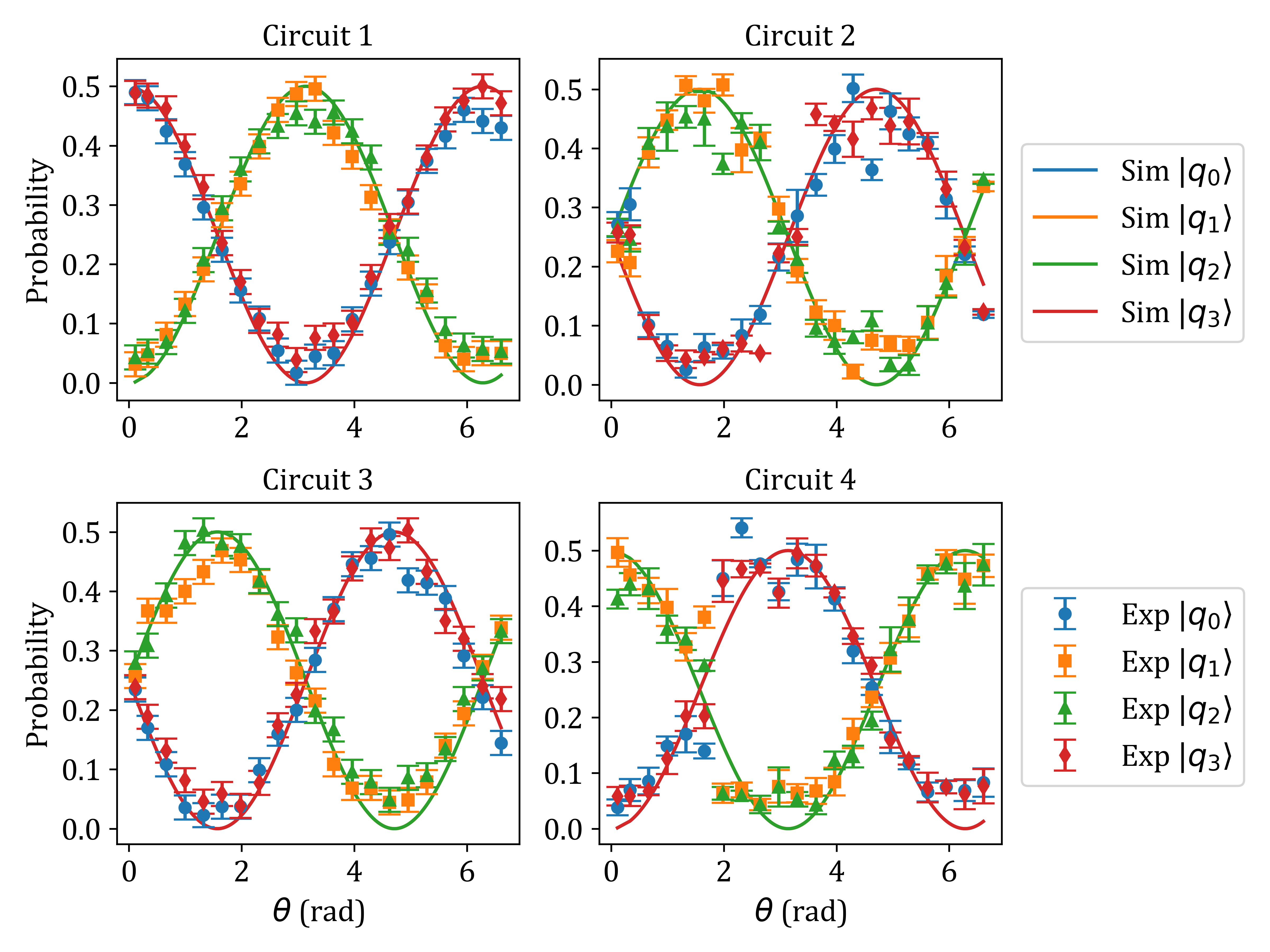}
  \caption{
  Measured state populations for the four effective CHSH measurement configurations as a function of the rotation angle \(\theta\). Experimental data are shown as points, while solid lines represent numerical simulations obtained from coherent evolution under a noiseless Hamiltonian model.
  }
  \label{CHSH2}
\end{figure}
\begin{figure}[t]
\centering
\includegraphics[width=\linewidth]{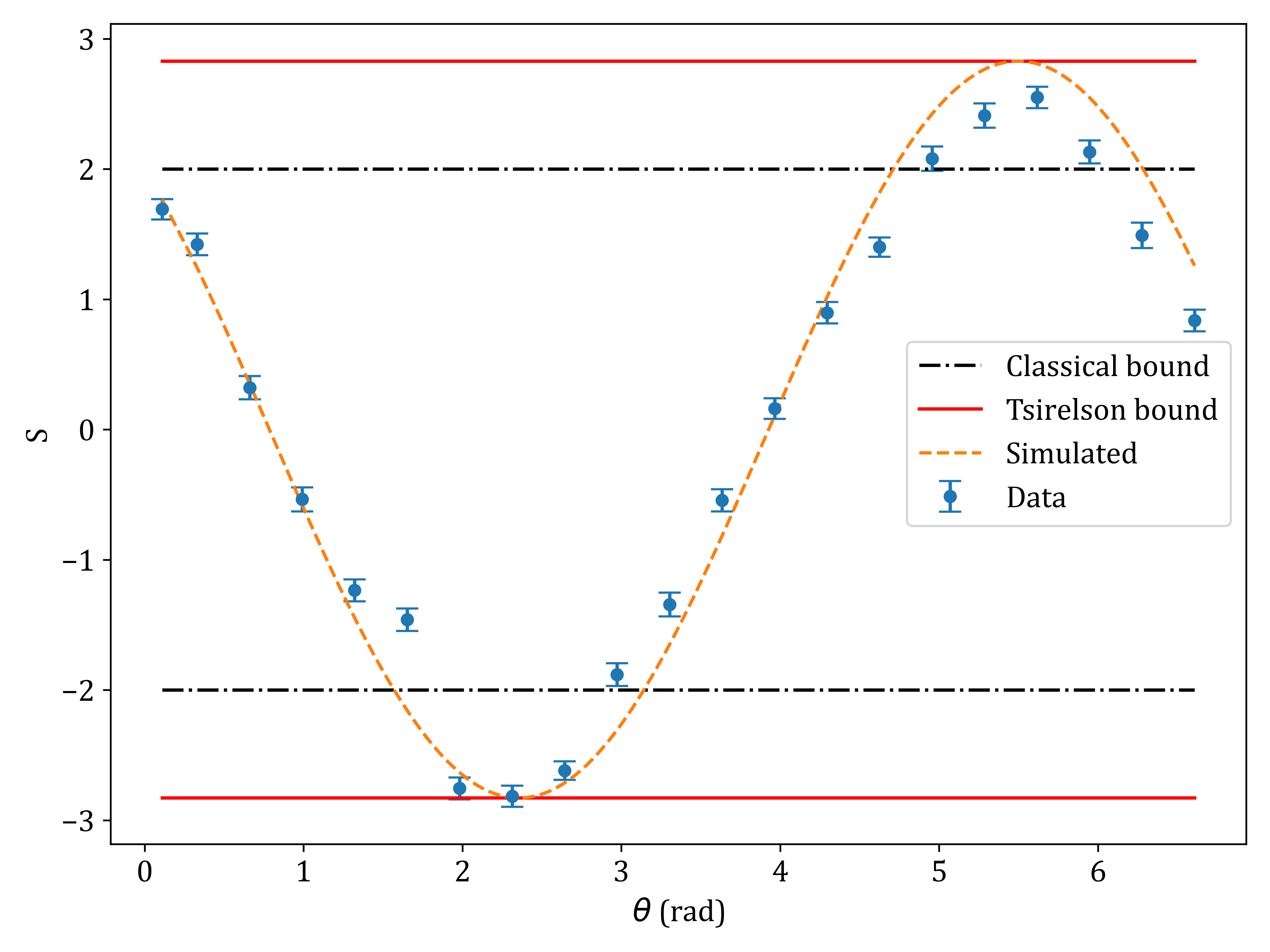}
\caption{ Measured CHSH parameter \(S\) as a function of the rotation angle \(\theta\). Experimental data are shown as points and numerical simulations as solid lines. The black horizontal line indicates the classical CHSH bound \(S=2\), while the red line denotes the Tsirelson bound \(S=2\sqrt{2}\). A maximum violation of \(S=2.816\pm0.082\) is observed near \(\theta = 3\pi/4\). The complete CHSH parameter \(S\) as a function of \(\theta\) is summarised in Table~II in the supplementary material.}
\label{CHSH3}
\end{figure}
\vspace{-5mm}
\subsection{State-dependent quantum contextuality}
To investigate nonclassical correlations in a single multilevel quantum system, we implement a CHSH-type contextuality protocol entirely within the four-dimensional Hilbert space. Under the qudit--qubit correspondence introduced above, the computational basis is mapped onto an effective two-qubit basis, allowing the standard CHSH observables to be realized using only coherent single-qudit operations. Since all measurements are performed on a single physical system, the experiment probes quantum contextuality rather than spatial Bell nonlocality.

Figure~\ref{CHSH1} illustrates the pulse sequences used to realize the four effective measurement settings,
$\langle ZZ\rangle$,
$\langle ZX\rangle$,
$\langle XZ\rangle$, and
$\langle XX\rangle$.
The initial sequence prepares the coherent state
\[
|\psi\rangle=\frac{|q_0\rangle+|q_3\rangle}{\sqrt2},
\]
which is isomorphic to the Bell state
$(|00\rangle+|11\rangle)/\sqrt2$
under the effective two-qubit mapping. Subsequent phase-controlled rotations define the required measurement bases before state-selective fluorescence detection. Together, these pulse sequences provide a complete implementation of the CHSH measurement protocol using only coherent operations within the four-level qudit.

For each rotation angle $\theta$, the experiment is repeated for all four measurement configurations. The measured state populations are shown in Fig.~\ref{CHSH2} together with numerical simulations based on coherent evolution under the experimentally calibrated Hamiltonian. The populations exhibit the expected coherent oscillations as the measurement basis is varied, and the experimental data closely follow the theoretical predictions over the full range of rotation angles. The remaining deviations are consistent with the finite state-preparation, coherent-control, and readout fidelities of the experiment, explained in the supplementary section.

The measured populations, $(P_{q_i})$, are converted into correlation functions according to
\begin{equation}
\langle AB\rangle
=
P_{q_0}-P_{q_1}-P_{q_2}+P_{q_3},
\quad
A,B\in\{X,Z\},
\end{equation}

from which the CHSH parameter is evaluated as
\begin{equation}
S=
\langle ZZ\rangle
-
\langle ZX\rangle
+
\langle XZ\rangle
+
\langle XX\rangle .
\label{eq:CHSH}
\end{equation}

Figure~\ref{CHSH3} presents the measured CHSH parameter as a function of the rotation angle together with the corresponding numerical simulation. The measured values closely reproduce the expected sinusoidal dependence predicted by coherent quantum evolution. A maximum violation of $S=2.816\pm0.082$ is obtained near $\theta=3\pi/4$, approaching the Tsirelson bound ($2\sqrt2\simeq2.828$) while clearly exceeding the classical noncontextual limit of $S=2$. A second local maximum is observed near $\theta=7\pi/4$, although with reduced amplitude because the longer pulse sequence accumulates additional decoherence and control errors during coherent evolution.

The observed violation demonstrates that strong contextual quantum correlations are preserved within a single programmable four-level trapped-ion qudit. 
\vspace{2mm}
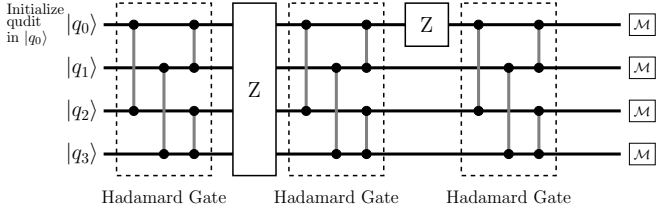
\begin{figure}[htbp]
\centering
\resizebox{\linewidth}{!}{%
\begin{tikzpicture}

    \node [font=\fontsize{10}{1}\selectfont, text width =  1.5cm] at (-1.5,3) {Initialize qudit in $|q_0\rangle$} ;

    \node [font=\fontsize{14}{1}\selectfont] at (-0.5,3) {$|q_0\rangle$} ; 
    \node [font=\fontsize{14}{1}\selectfont] at (-0.5,2) {$|q_1\rangle$} ; 
    \node [font=\fontsize{14}{1}\selectfont] at (-0.5,1) {$|q_2\rangle$} ; 
    \node [font=\fontsize{14}{1}\selectfont] at (-0.5,0) {$|q_3\rangle$} ; 
    
    \draw[black, line width = 2pt] (0,3) -- (12,3);
    \draw[black, line width = 2pt] (0,2) -- (12,2);
    \draw[black, line width = 2pt] (0,1) -- (12,1);
    \draw[black, line width = 2pt] (0,0) -- (12,0);

    \draw[gray, line width = 2pt] (0.7,3) -- (0.7,1);
    \draw[gray, line width = 2pt] (1.4,2) -- (1.4,0);
    \draw[gray, line width = 2pt] (2.1,3) -- (2.1,2);
    \draw[gray, line width = 2pt] (2.1,1) -- (2.1,0);
    \filldraw (0.7,3) circle (3pt); \filldraw (0.7,1) circle (3pt);
    \filldraw (1.4,2) circle (3pt); \filldraw (1.4,0) circle (3pt);
    \filldraw (2.1,3) circle (3pt); \filldraw (2.1,2) circle (3pt);
    \filldraw (2.1,1) circle (3pt); \filldraw (2.1,0) circle (3pt);
    \draw [dashed,line width = 1pt] (0.3,3.5) rectangle (2.5,-0.5);
    \node [font=\fontsize{12}{1}\selectfont] at (1.4,-1) {Hadamard Gate} ; 

    \draw [fill=white!30, line width = 1pt] (3,3.5) rectangle (4,-0.5);
    \node [font=\fontsize{14}{1}\selectfont] at (3.5,1.5) {Z} ; 

    \draw[gray, line width = 2pt] (4.7,3) -- (4.7,1);
    \draw[gray, line width = 2pt] (5.4,2) -- (5.4,0);
    \draw[gray, line width = 2pt] (6.1,3) -- (6.1,2);
    \draw[gray, line width = 2pt] (6.1,1) -- (6.1,0);
    \filldraw (4.7,3) circle (3pt); \filldraw (4.7,1) circle (3pt);
    \filldraw (5.4,2) circle (3pt); \filldraw (5.4,0) circle (3pt);
    \filldraw (6.1,3) circle (3pt); \filldraw (6.1,2) circle (3pt);
    \filldraw (6.1,1) circle (3pt); \filldraw (6.1,0) circle (3pt);
    \draw [dashed,line width = 1pt] (4.3,3.5) rectangle (6.5,-0.5);
    \node [font=\fontsize{12}{1}\selectfont] at (5.4,-1) {Hadamard Gate} ;

    \draw [fill=white!30, line width = 1pt] (7,3.5) rectangle (8,2.5);
    \node [font=\fontsize{14}{1}\selectfont] at (7.5,3) {Z} ; 

    \draw[gray, line width = 2pt] (8.7,3) -- (8.7,1);
    \draw[gray, line width = 2pt] (9.4,2) -- (9.4,0);
    \draw[gray, line width = 2pt] (10.1,3) -- (10.1,2);
    \draw[gray, line width = 2pt] (10.1,1) -- (10.1,0);
    \filldraw (8.7,3) circle (3pt); \filldraw (8.7,1) circle (3pt);
    \filldraw (9.4,2) circle (3pt); \filldraw (9.4,0) circle (3pt);
    \filldraw (10.1,3) circle (3pt); \filldraw (10.1,2) circle (3pt);
    \filldraw (10.1,1) circle (3pt); \filldraw (10.1,0) circle (3pt);
    \draw [dashed,line width = 1pt] (8.3,3.5) rectangle (10.5,-0.5);
    \node [font=\fontsize{12}{1}\selectfont] at (9.4,-1) {Hadamard Gate} ;
    
    \node[meas] at (12.5,3) {$\mathcal{M}$};
    \node[meas] at (12.5,2) {$\mathcal{M}$};
    \node[meas] at (12.5,1) {$\mathcal{M}$};
    \node[meas] at (12.5,0) {$\mathcal{M}$};
\end{tikzpicture}%
}
\caption{
Quantum circuit implementing Grover’s search algorithm in a single \(d=4\) qudit. Connected pairs of levels denote coherent rotations within the corresponding two-level subspaces, forming embedded \(SU(2)\) operations inside the full \(SU(4)\) manifold. The initial Hadamard-like transformation prepares an equal superposition of all four computational basis states. The central multi-level \(Z\)-operation acts as the Grover oracle by selectively applying a \(\pi\)-phase shift to the marked state, thereby inverting its phase relative to the remaining basis states. The subsequent Hadamard--\(Z\)--Hadamard sequence implements the Grover diffusion operator \(U_s = 2|s\rangle\langle s| - I\), corresponding to reflection about the mean amplitude in Hilbert space.
}
\label{GroverCircuit}
\end{figure}
\begin{figure}[t]
    \centering

    \includegraphics[width=0.9\columnwidth]{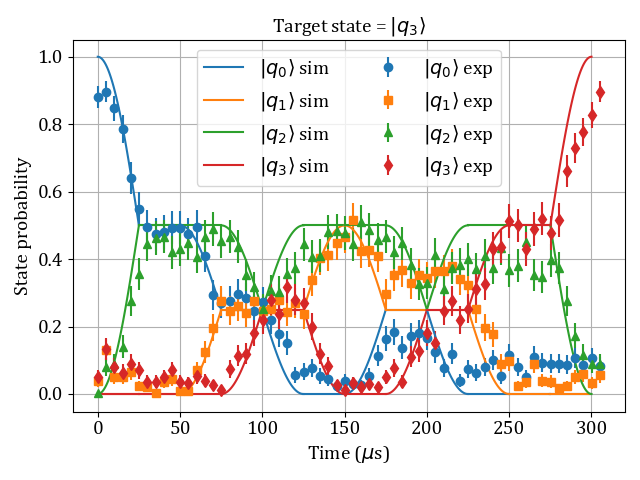}

    \vspace{0mm}

    \includegraphics[width=0.9\columnwidth]{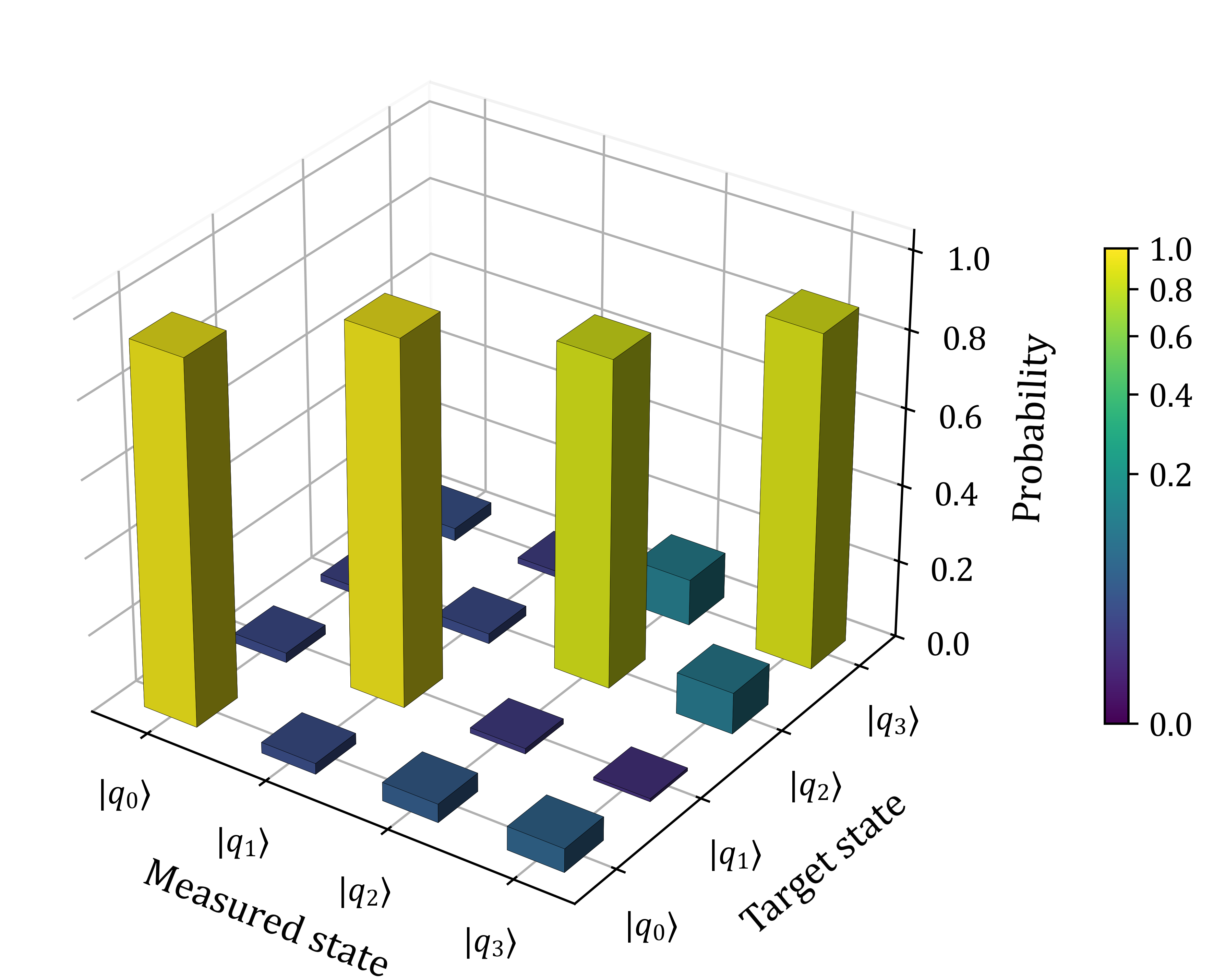}

    \caption{
    Experimental realization of Grover's search in a four-dimensional trapped-ion qudit.
    (Top:) Evolution of the computational basis-state populations during Grover's algorithm for the marked state $|q_3\rangle$. Symbols represent experimental data, while solid curves are QuTiP simulations.
    (Bottom:) Measured output-state probability matrix for all four target states. The dominant diagonal elements confirm successful retrieval of the marked state, with search fidelities of $(94.2\pm1.6)\%$, $(94.5\pm2.0)\%$, $(84.8\pm3.3)\%$, and $(86.8\pm3.2)\%$ for target states $|q_0\rangle$, $|q_1\rangle$, $|q_2\rangle$, and $|q_3\rangle$, respectively. The complete probability matrix and statistical uncertainties are provided in the Supplementary Material.
    }
    \label{fig:grover_results}
\end{figure}

\subsection{Entanglement-free Grover search}
Having established universal coherent control of the four-dimensional Hilbert space, we next implement Grover's quantum search algorithm. Under the qudit--qubit correspondence introduced in Eq.~(\ref{eq:qubit_mapping}), the four computational basis states emulate an effective two-qubit register while remaining encoded in a single physical quantum system. Consequently, the search algorithm is implemented through coherent interference within one multilevel wavefunction rather than entanglement between physically distinct qubits, following the framework proposed by Lloyd~\cite{Lloyd1999} and later analyzed by Meyer~\cite{Meyer2000}.

Figure~\ref{GroverCircuit} illustrates the effective Grover circuit implemented in the present work. In contrast to the conventional multi-qubit circuit architecture discussed in Ref.~\cite{Meyer2000}, where Grover iterations are represented using separate qubit lines and repeated oracle--diffusion blocks, the present implementation embeds the entire algorithm directly within the \(SU(4)\) manifold. In the original qubit-based formulation of Grover search, the algorithm is implemented using tensor-product operations acting on physically distinct qubits together with entangling oracle operations. By comparison, the present architecture realizes the complete search dynamics through coherent rotations between selected subspaces of a single four-level Hilbert space. The required oracle and diffusion operations are synthesized using embedded \(SU(2)\) rotations that collectively generate the effective \(SU(4)\) evolution.

As explained in Fig.~\ref{GroverCircuit}, the ion is initialized in the computational basis state $|q_0\rangle$. The first Hadamard block (left dashed box) consists of a sequence of embedded $SU(2)$ rotations that prepares the equal superposition
\begin{equation}
|s\rangle
=
\frac{1}{2}
\sum_{i=0}^{3}|q_i\rangle,
\label{eq:grover_superposition}
\end{equation}
distributing the probability amplitude uniformly among the four computational basis states.

The central multilevel phase gate acts as the Grover oracle by selectively applying a $\pi$ phase shift to the marked state while leaving the remaining basis states unchanged. Since only the relative phase is modified, the state populations remain unchanged immediately after the oracle.

The final Hadamard--$Z$--Hadamard sequence (right dashed box) implements the diffusion operator, which reflects the quantum state about the average probability amplitude,
\begin{equation}
U_s
=
2|s\rangle\langle s|-I.
\label{eq:diffusion}
\end{equation}
This operation converts the phase difference introduced by the oracle into constructive interference for the marked state and destructive interference for all remaining states, thereby amplifying the probability of measuring the desired solution after a single Grover iteration.

To characterize the coherent search dynamics, the pulse sequence was interrupted at different stages and the instantaneous state populations were measured using state-selective fluorescence detection. Figure~\ref{fig:grover_results}(top) shows the measured population evolution for a search targeting the state $|q_3\rangle$, together with numerical simulations obtained using the experimentally calibrated Hamiltonian. Following initialization, the first Hadamard operation prepares an approximately uniform superposition of all four basis states. The subsequent oracle leaves the populations unchanged while introducing the required relative phase, after which the diffusion operator transfers population coherently toward the marked state through quantum interference. The measured dynamics closely follow the theoretical prediction throughout the complete pulse sequence.

The search protocol was subsequently repeated for each computational basis state. Figure~\ref{fig:grover_results}(bottom) summarizes the measured target-state success probabilities after one Grover iteration. Success probabilities exceeding $84.8\%$ are obtained for all four target states, reaching a maximum of $94.5\%$ with an average success probability of $90.1\%$. The remaining deviations from the ideal result are dominated by coherent control imperfections and state-detection errors explained in detail in the supplementary section. 

Compared with conventional two-qubit implementations, the single-qudit realization exhibits substantially improved performance over previously reported trapped-ion demonstrations, where search fidelities of approximately \(60\%\) were observed~\cite{Brickman2005}, and achieves fidelities comparable to superconducting-qubit implementations reporting fidelities exceeding \(89\%\) (and up to \(98\%\) with error correction)~\cite{Pokharel2024}. The enhanced performance observed in the present architecture is likely attributable to the suppression of common-mode systematic errors within a single coherent quantum system, in contrast to implementations relying on interactions between physically distinct qubits.

\section{Discussion}

This work establishes a programmable trapped-ion optical qudit as a unified platform for quantum information processing and quantum foundations. By combining universal coherent $SU(4)$ control with high-fidelity optical manipulation in a single $^{138}\mathrm{Ba}^{+}$ ion, we demonstrate both an entanglement-free Grover search and a CHSH-type contextuality test within the same four-dimensional Hilbert space. The target-state success probabilities reach up to $94.5\pm2.0\%$, while the maximum contextuality violation, $S=2.816\pm0.082$, closely approaches the Tsirelson bound. The main limitations arise from laser phase noise, state-detection errors, finite counting statistics, and single-qudit gate imperfections, with slow laser-frequency and magnetic-field fluctuations limiting long-term coherence. A detailed error analysis is provided in the Supplementary Information.

Unlike implementations based on entangling gates between separate subsystems, the Grover search operates entirely within a single multilevel quantum system, with the search enhancement arising from coherent interference generated through programmable $SU(4)$ dynamics. This does not imply that multipartite entanglement is unnecessary for scalable quantum speedups; rather, it demonstrates how qudits can exploit higher-dimensional Hilbert spaces while reducing physical carrier overhead. The observed contextuality violation further shows that the same control architecture supports strong non-classical correlations.

The combination of coherent control, algorithmic benchmarking, and contextuality measurements provides a platform for investigating the relationship between quantum interference, contextuality, and computational performance, complementing theoretical proposals identifying contextuality as a resource for universal and magic-state quantum computation~\cite{Howard2014}. More broadly, the results highlight the potential of trapped-ion qudits as resource-efficient quantum processors that exploit atomic multilevel structure within a single physical carrier.

Looking forward, these findings motivate two open questions: (1) whether resource-efficient qudit-based algorithms exist that can demonstrate quantum advantage by relying explicitly on contextuality as a resource, and (2) whether joint contextuality and algorithmic-performance benchmarking of the kind demonstrated here can provide a useful metric for assessing the computational capability and scalability of programmable high-dimensional quantum processors.
%
\section{Method}
Experiments are performed using a single trapped $^{138}\mathrm{Ba}^{+}$ ion confined in a linear Paul trap~\cite{DuttaClassifier}. A four-dimensional optical qudit is encoded in two Zeeman sublevels of the $6S_{1/2}$ ground-state and two of the metastable $5D_{5/2}$ manifold. A static magnetic field defines the quantization axis and lifts the Zeeman degeneracy, enabling frequency-selective optical addressing. Each experimental cycle comprises Doppler cooling, optical pumping, coherent manipulation, and state-selective fluorescence detection, with experimental uncertainties detailed in the Supplementary Information.

Programmable coherent control is realized using phase-coherent optical fields synthesized by an arbitrary waveform generator, providing independent control of amplitude, frequency, and phase. Resonant driving between qudit states implements $SU(2)$ rotations, while concatenated rotations synthesize multilevel operations. Under the mapping to an effective two-qubit basis, the transition pairs $\{R_{01},R_{23}\}$ and $\{R_{02},R_{13}\}$ implement logical single-qubit operations and collectively provide universal $SU(4)$ control. This framework enables both Grover search and CHSH contextuality circuits through programmable composite pulse sequences without inter-particle entangling gates.\\
\subsection{Mathematical framework for coherent \texorpdfstring{$SU(4)$}{SU(4)} control}
Universal control of the four-dimensional optical qudit is achieved through resonant optical rotations acting on experimentally accessible two-level subspaces. An elementary rotation is described by
\begin{equation}
R_{ij}(\theta,\phi)=
\exp\left[
-\frac{i\theta}{2}
\left(
e^{-i\phi}|i\rangle\langle j|
+
e^{i\phi}|j\rangle\langle i|
\right)
\right],
\label{eq:Rij}
\end{equation}
where $\theta=\Omega t$ is the pulse area and $\phi$ is the optical phase. The experimentally accessible rotations $\{R_{01},R_{02},R_{13},R_{23}\}$ generate the full Lie algebra ${SU}(4)$, enabling arbitrary unitary operations within the four-dimensional Hilbert space.

The computational basis presented in eq.~\ref{eq:qubit_mapping}
establishes an isomorphism between the native qudit and an effective two-qubit system. Consequently, any logical single-qubit operation
\vspace{-5mm}
\begin{equation}
G=
\begin{pmatrix}
a & b \\
c & d
\end{pmatrix}
\in SU(2),
\qquad
a,b,c,d \in \mathbb{C},
\end{equation}
denote an arbitrary single-qubit unitary operation satisfying
$G^\dagger G=\mathbb{I},
\quad
\det(G)=1$,
is embedded into the four-dimensional space as
\begin{equation}
U_A=
\begin{pmatrix}
a&0&b&0\\
0&a&0&b\\
c&0&d&0\\
0&c&0&d
\end{pmatrix},
\qquad
U_B=
\begin{pmatrix}
a&b&0&0\\
c&d&0&0\\
0&0&a&b\\
0&0&c&d
\end{pmatrix},
\end{equation}
which are experimentally synthesized using the native rotations,
\[
U_A=R_{02}R_{13},
\qquad
U_B=R_{01}R_{23},
\] up to programmable pulse areas and optical phases. This framework provides the basis for implementing both the Grover search algorithm and the contextuality measurements described below.

\subsection{Implementation of the Grover search algorithm}
The Grover search circuit was synthesized entirely from the experimentally accessible transition-selective rotations described in the previous section. The generalized Hadamard operation was implemented as a composite sequence of embedded $SU(2)$ rotations that prepares the equal superposition
\vspace{-5mm}
\begin{equation}
|s\rangle
=
U_H|q_0\rangle
=
\frac12
\sum_{i=0}^{3}|q_i\rangle .
\end{equation}
The oracle operation applies a selective $\pi$ phase shift to the marked computational basis state,
\begin{equation}
U_f|x\rangle
=
(-1)^{f(x)}|x\rangle,
\end{equation}
where $f(x)=1$ for the marked state and $0$ otherwise.

The diffusion operator is given by
\begin{equation}
U_s
=
2|s\rangle\langle s|-I
=
U_H
\left(
2|q_0\rangle\langle q_0|-I
\right)
U_H^\dagger ,
\end{equation}

and is realized experimentally through the second Hadamard--$Z$--Hadamard sequence shown in Fig.~\ref{GroverCircuit}. The complete Grover iteration is $ U_G
=
U_sU_f .$

Each unitary was decomposed into sequences of the experimentally available transition-selective operations
\(R_{01}\),
\(R_{02}\),
\(R_{13}\), and
\(R_{23}\).
The pulse phases, amplitudes and durations were programmed using the arbitrary waveform generator to generate the required composite $SU(4)$ evolution. Numerical simulations were performed using QuTiP with the experimentally calibrated Hamiltonian and measured Rabi frequencies.
\begin{figure}
\centering
\includegraphics[width=\linewidth]{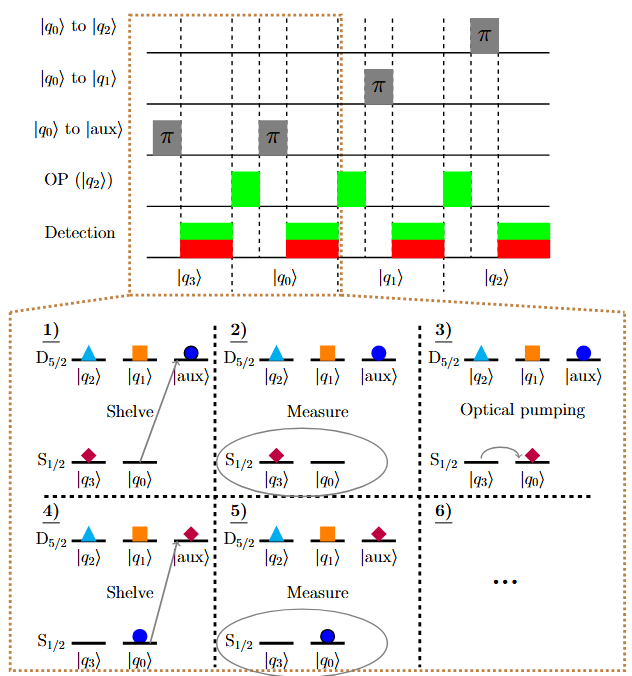}
\caption{Experimental sequence for qudit state readout: The populations of the four computational basis states are measured through a sequence of state-selective shelving, optical pumping, and fluorescence detection operations. In each detection stage, the population of the selected qudit state is mapped onto the reference state and measured using shelving-assisted fluorescence detection, enabling reconstruction of the complete qudit-state population distribution. A detailed description of the readout protocol is provided in Section~\ref{sec:readout}.
 }
\label{fig: state-measurement-pulse}
\end{figure}
\vspace{-5mm}

\subsection{Qudit state readout and uncertainties}
\label{sec:readout}
State detection is performed using shelving-based fluorescence readout, as illustrated in Fig.~\ref{fig: state-measurement-pulse}. A single reference state, $|q_0\rangle$, and an auxiliary shelving state (Fig.~\ref{qudit}) are used to determine the populations of all four computational states. The measured state-detection fidelity is $98.6\%$~\cite{DuttaClassifier,Yum2017BaQubit}. 
The readout sequence is summarized as follows:\\
$|q_3\rangle$ population: The population in $|q_0\rangle$ is shelved to the auxiliary state using $R_{q_0\text{-aux}}(\pi,0)$. Fluorescence detection on the $6S_{1/2}\leftrightarrow6P_{1/2}$ transition then measures the unshelved population, providing the $|q_3\rangle$ population.\\
$|q_0\rangle$ population: The ion is optically pumped back to $|q_0\rangle$, followed by shelving with $R_{q_0\text{-aux}}(\pi,0)$ and fluorescence detection to determine the $|q_0\rangle$ population.\\
$|q_2\rangle$ population: After reinitialization to $|q_0\rangle$, a resonant swap between $|q_0\rangle$ and $|q_2\rangle$ maps the $|q_2\rangle$ population onto $|q_0\rangle$. The same shelving and fluorescence sequence is then applied.\\
$|q_1\rangle$ population: Similarly, a resonant $|q_0\rangle\leftrightarrow|q_1\rangle$ swap maps the $|q_1\rangle$ population onto $|q_0\rangle$, followed by shelving and fluorescence detection.\\

Thus, the complete population distribution is obtained by changing only the final mapping operation before the common shelving and fluorescence sequence. This procedure reduces systematic drifts and avoids additional errors associated with separate measurement protocols.

The dominant uncertainties arise from state detection, coherent control, and slow laser and magnetic-field fluctuations. Optical pumping prepares $|q_0\rangle$ with approximately $99.8\%$ fidelity, while the single-qudit $\pi$-pulse fidelity is $(98.8\pm0.3)\%$, determined from Rabi oscillations (Fig.S2). The remaining gate error is mainly attributed to pulse-area imperfections and weak off-resonant excitation. 

The selected transition geometry, using transitions with opposite or nearly symmetric Zeeman sensitivities, suppresses first-order magnetic-field dephasing. Addressing only the relevant two-level subspace also reduces differential AC Stark shifts and spectator-state coupling. Two-stage Doppler cooling yields $\bar{n}\sim8$, sufficient for operation in the Lamb--Dicke regime, while Rabi and Ramsey calibrations compensate laser-frequency and phase drifts. A quantitative breakdown of the dominant experimental uncertainties is summarized in Table~I in the supplementary section. Additional characterization of gate performance, noise analysis, and long-term stability measurements is provided in the Supplementary Information.
\bibliography{bibliography}
\section*{Supplementary Information}

\setcounter{equation}{0}
\setcounter{figure}{0}
\setcounter{table}{0}
\setcounter{page}{1}

\renewcommand{\theequation}{S\arabic{equation}}
\renewcommand{\thefigure}{S\arabic{figure}}
\renewcommand{\thetable}{S\arabic{table}}
\renewcommand{\citenumfont}[1]{S#1}
\renewcommand{\bibnumfmt}[1]{[S#1]}
\makeatother
\section{Experimental uncertainties}

The performance of the optical qudit is primarily limited by laser-induced decoherence, coherent control errors, and state-detection infidelity. While these error sources are broadly similar to those encountered in trapped-ion optical qubits~\cite{Yum2017BaQubit, DuttaClassifier, Dutta2020NoiseProbe}, multilevel coherent control introduces additional sensitivity to differential phase accumulation and off-resonant coupling between nearby transitions.

To improve robustness against magnetic-field noise, the qudit is encoded using transitions with opposite Zeeman sensitivities, predominantly \(\Delta m=0\) and \(\Delta m=\pm1\). The resulting partially compensating frequency shifts suppress first-order magnetic dephasing during sequential gate operations. The selected transition configuration also reduces differential AC Stark shifts by limiting each operation to only two relevant optical transitions.

State-preparation errors arise mainly from imperfect optical pumping and residual polarization impurities. Since initialization in \(|q_0\rangle\) exhibits the highest fidelity, all experimental sequences begin from this state. Residual magnetic-field drift is further suppressed by a stable permanent-magnet quantization field.

Laser-induced decoherence constitutes the dominant contribution to the overall error budget owing to the narrow optical transitions used for coherent control. The principal limitations originate from laser-frequency drift, phase noise, pulse-area fluctuations, and residual AC Stark shifts. These effects are mitigated by balancing the Rabi frequencies of all addressed transitions and performing regular Rabi and Ramsey calibrations to compensate slow frequency and phase drifts.

\begin{figure*}[t]
\centering
\subfloat[\label{fig:ramsey-coherence}]{%
    \includegraphics[width=0.48\textwidth]{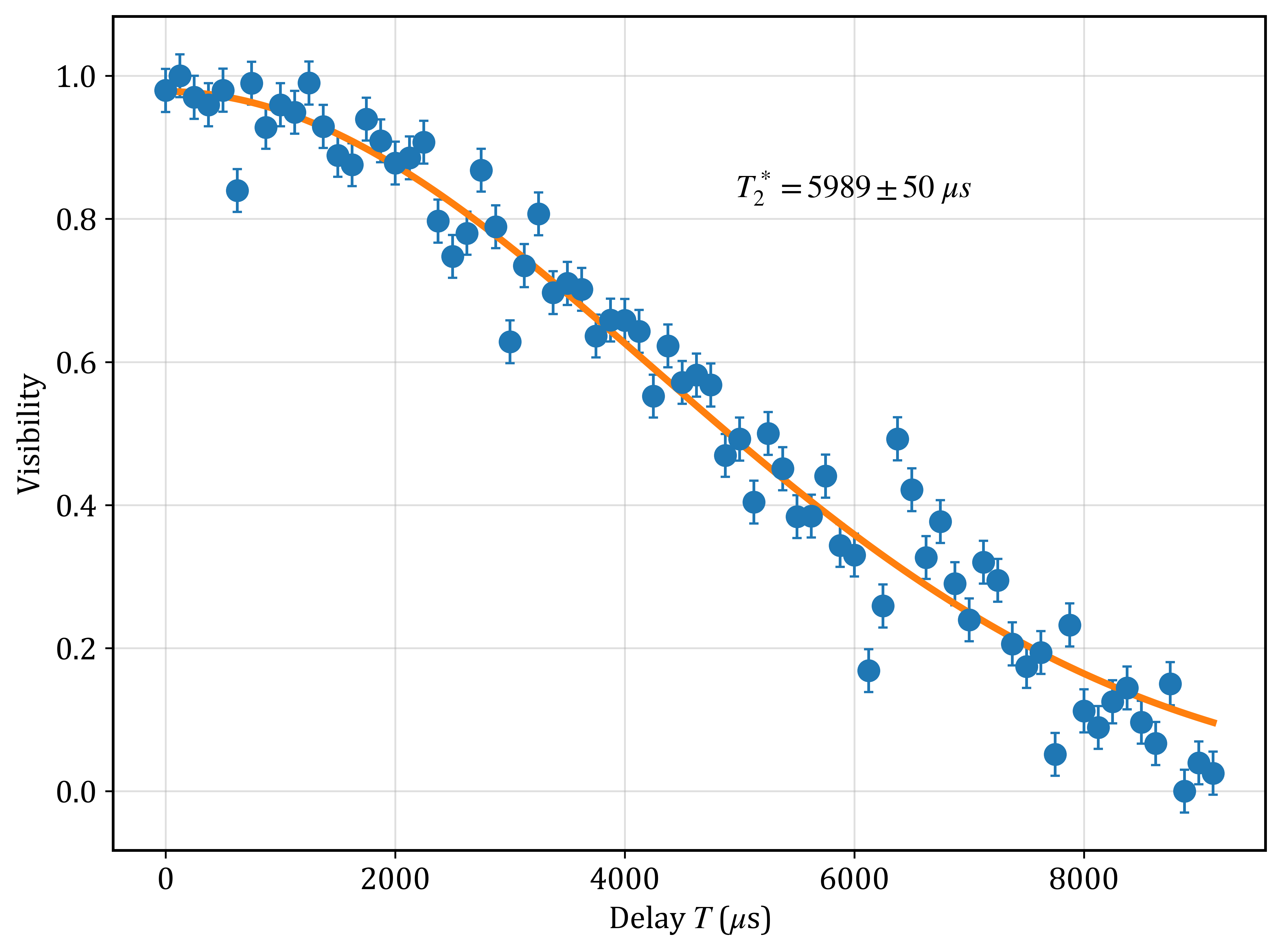}%
}
\hfill
\subfloat[\label{fig:qudit-gate}]{%
    \includegraphics[width=0.48\textwidth]{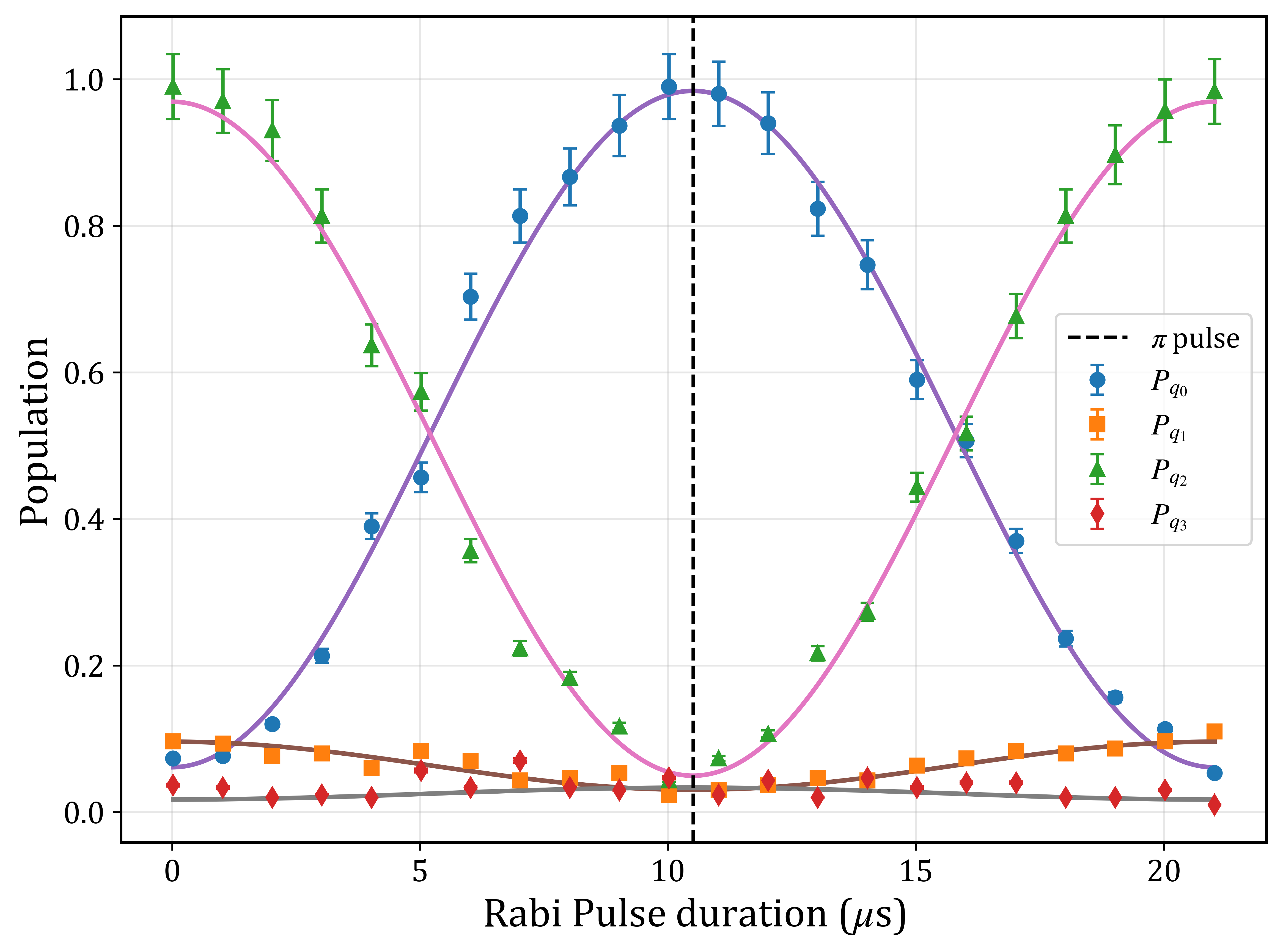}%
}
\caption{Characterization of coherent control in the $d=4$ optical-qudit manifold.
(a) Gaussian fit to the Ramsey fringe envelope obtained from the measured visibility as a function of Ramsey delay time. The solid curve represents the fit to $V(T)=V_0\exp[-(T/T_2^*)^2]$, yielding the optical coherence time $T_2^*$ and the initial visibility $V_0$. Error bars denote one standard deviation obtained from repeated measurements. (b) Coherent Rabi oscillations between the coupled qudit states. The measured normalized populations of the computational basis states $\{|q_0\rangle, |q_1\rangle, |q_2\rangle, |q_3\rangle\}$ are shown as a function of the resonant pulse duration for the single-qudit rotation $R_{02}(\theta,\phi)$. Symbols with one-standard-deviation error bars represent the experimentally measured state populations, while the solid curves are simultaneous Rabi fits sharing a common $\pi$-pulse duration. Coherent population transfer is observed predominantly between the coupled states $|q_0\rangle$ and $|q_2\rangle$, whereas the spectator states $|q_1\rangle$ and $|q_3\rangle$ remain nearly unpopulated throughout the evolution, confirming selective and phase-coherent control of the four-level qudit. The vertical dashed line denotes the fitted resonant $\pi$-pulse duration corresponding to the maximum population transfer from $|q_0\rangle$ to $|q_2\rangle$. The normalized state populations at the resonant $\pi$ pulse provide the gate characterization, yielding a single-qudit $\pi$-pulse fidelity of $(98.8 \pm 0.3)\%$.}
\label{fig:coherence-gate}
\end{figure*}

Additional control errors arise from the sequential application of multiple optical frequencies, where imperfect synchronization and weak off-resonant excitation of neighboring transitions reduce gate fidelity. Nevertheless, coherent Rabi measurements (Fig.~\ref{fig:qudit-gate}) yield a single-qudit \(\pi\)-pulse fidelity of \((98.8 \pm 0.3)\%\), demonstrating high-fidelity coherent control.

State-detection fidelity is primarily limited by imperfect shelving transfer, photon shot noise, and overlap of fluorescence photon-count distributions~\cite{Yum2017BaQubit}, whereas spontaneous decay from the metastable \(5D_{5/2}\) manifold and slow environmental fluctuations contribute only minor errors over the experimental timescale. The dominant experimentally observed error sources are summarized in Table~\ref{tab:qudit_errors}.

\begin{table}[htbp]
\centering
\caption{Estimated uncertainty budget for the measured CHSH parameter. Relative contributions are inferred from experimentally characterized system performance described in the text. Additional systematic effects, including laser-intensity noise, differential AC Stark shifts, AOM efficiency drift, state leakage, beam alignment drift, long-term drift, correlated 1/f noise and shelving errors, are included in the overall uncertainty estimate but are not listed individually owing to their comparatively small contributions.}
\label{tab:qudit_errors}
\small
\begin{tabular}{lc}
\hline
\hline
Error source & Relative contribution (\%) \\
\hline
Finite counting statistics & 1.56 \\
State preparation & 0.20 \\
State detection & 1.40 \\
Single-qudit gate fidelity & 1.20 \\
Laser-frequency drift & 0.50 \\
Magnetic-field fluctuations & 0.40 \\
Off-resonant excitation & 0.20 \\
\hline
\hline
\end{tabular}
\end{table}

\subsection{Coherent qudit operations}

All coherent operations are generated from phase-coherent RF tones produced by an arbitrary waveform generator. The amplitudes of the driving fields are individually adjusted so that all addressed transitions exhibit a common $\pi$-pulse duration. Prior to each experimental run, resonance frequencies and pulse phases are calibrated using Rabi and Ramsey measurements to compensate slow laser frequency drift and phase evolution.

Figure~\ref{fig:coherence-gate} summarizes the calibration measurements, showing representative Rabi oscillations together with Ramsey coherence data used to determine the resonance conditions and coherence properties of the selected qudit transitions.

The complete Grover-search and CHSH pulse sequences are implemented through phase-coherent concatenation of these calibrated elementary rotations. Each experimental cycle, including preparation, manipulation, and measurement, requires approximately $8$--$10~\mathrm{ms}$, substantially shorter than the lifetime of the $5D_{5/2}$ manifold.
\begin{figure}[t]
\centering
\includegraphics[width=\linewidth]{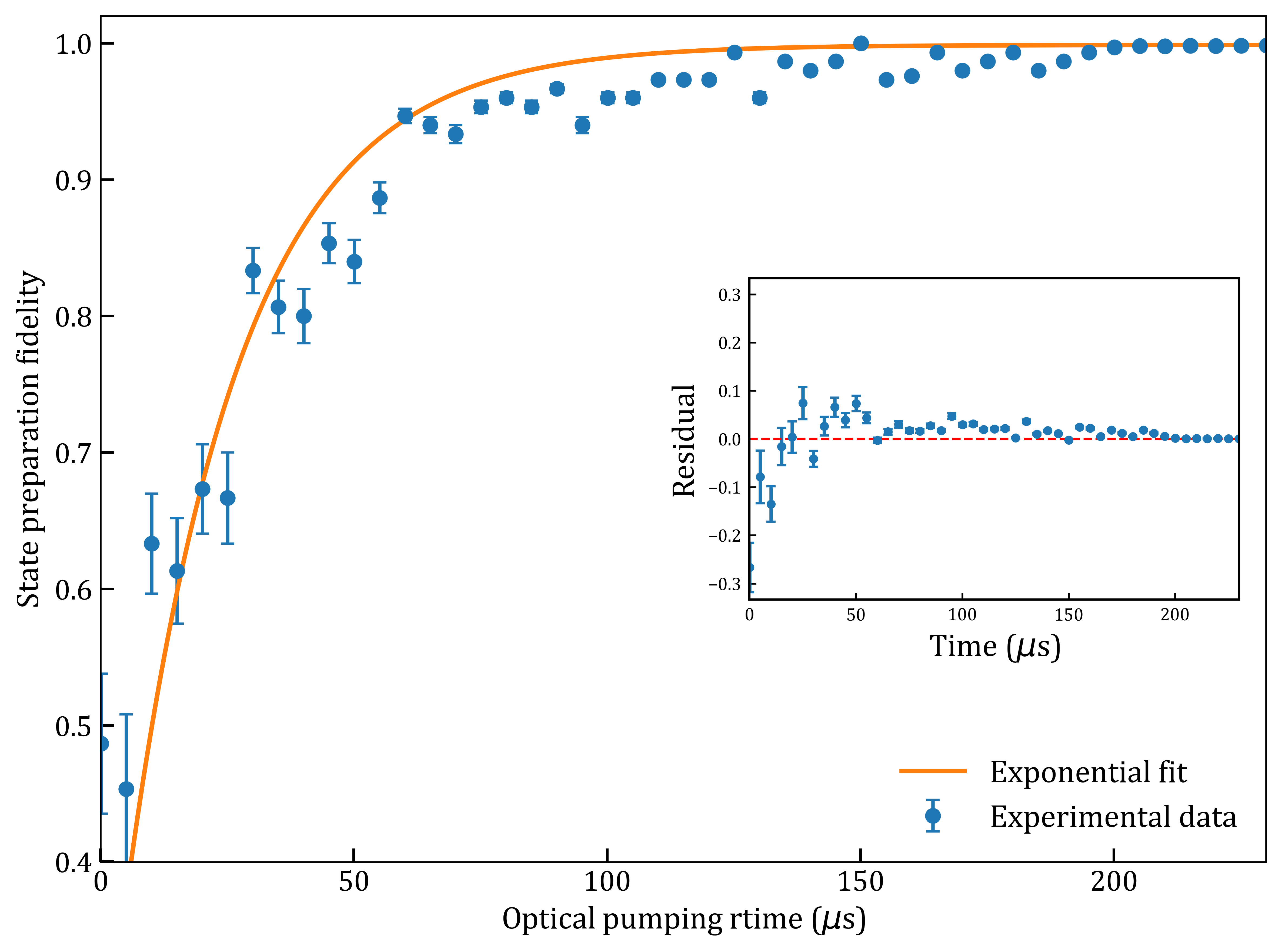}
\caption{State-preparation fidelity as a function of optical pumping duration. Experimental data are shown as discrete points with statistical error bars, while the solid curve represents a weighted exponential fit to the optical pumping dynamics. The fidelity increases monotonically with pumping duration and saturates near unity at long interaction times, demonstrating efficient population transfer into the target state. The residual population is suppressed to the \(10^{-3}\) level, corresponding to a state-preparation fidelity approaching \(99.8\%\). (Inset) The residuals were calculated from the difference between the experimentally measured populations and the fitted exponential model. Although the statistical uncertainties remain comparatively small, the residual fluctuations exceed the corresponding error bars for several data points, indicating that systematic effects dominate the experimental uncertainty.}
\label{fig:fidelityfit}
\end{figure}
\begin{figure*}[t]
\centering
\subfloat[]{%
\includegraphics[width=0.42\linewidth]{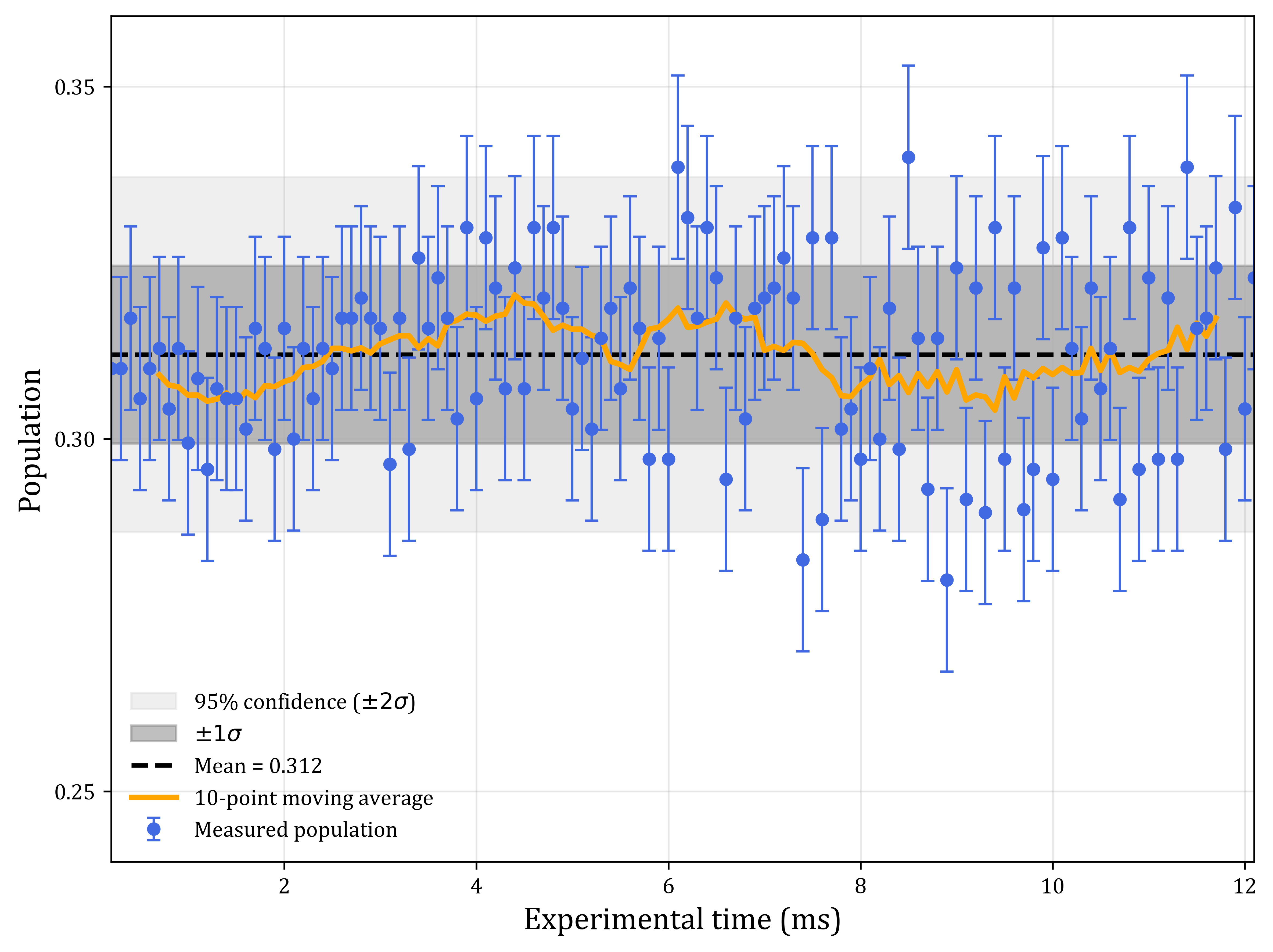}%
\label{fig:population}}
\hfill
\subfloat[]{%
\includegraphics[width=0.42\linewidth]{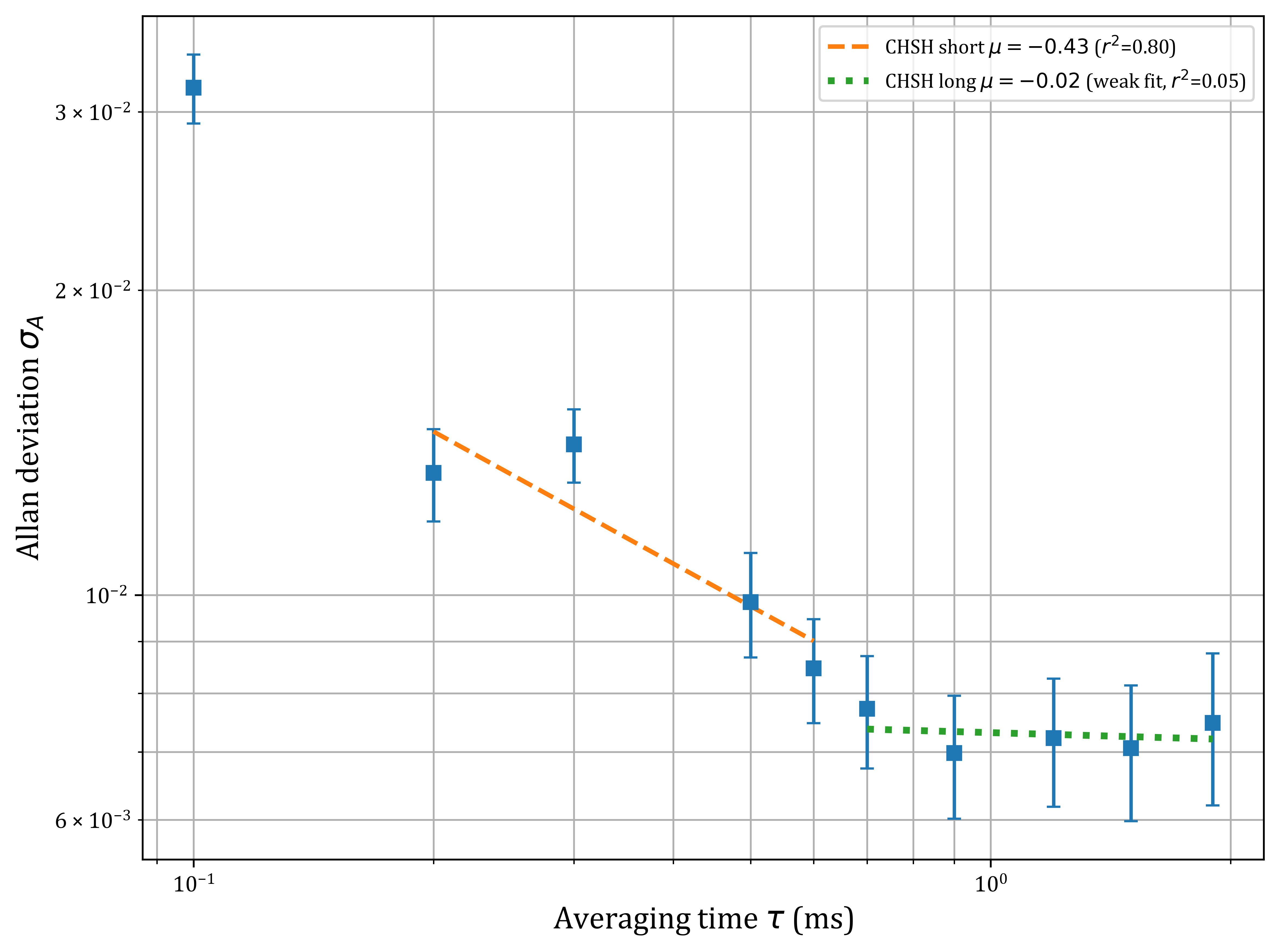}%
\label{fig:allan}}

\vspace{0.3cm}

\subfloat[]{%
\includegraphics[width=0.42\linewidth]{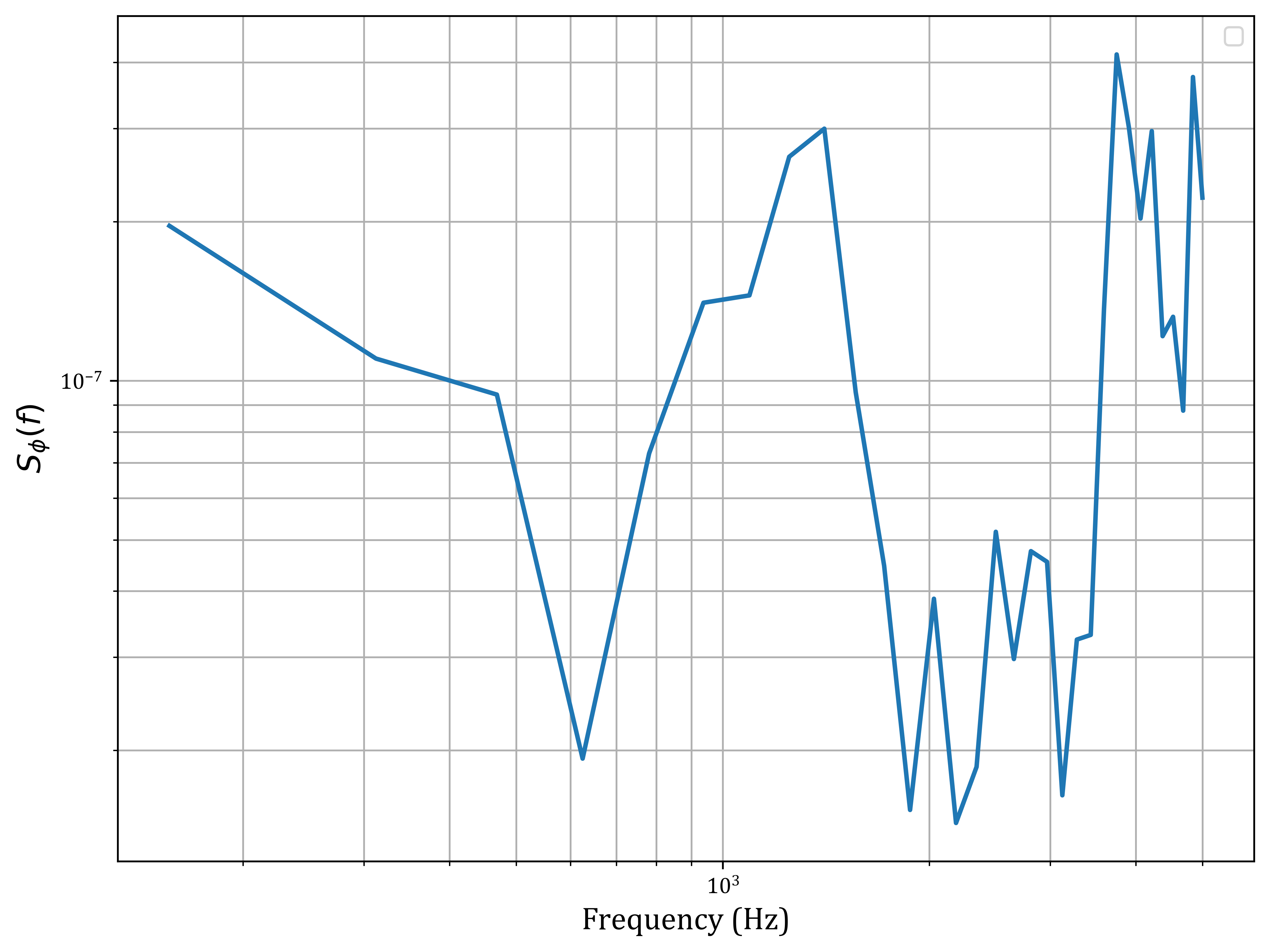}%
\label{fig:psd}}
\hfill
\subfloat[]{%
\includegraphics[width=0.42\linewidth]{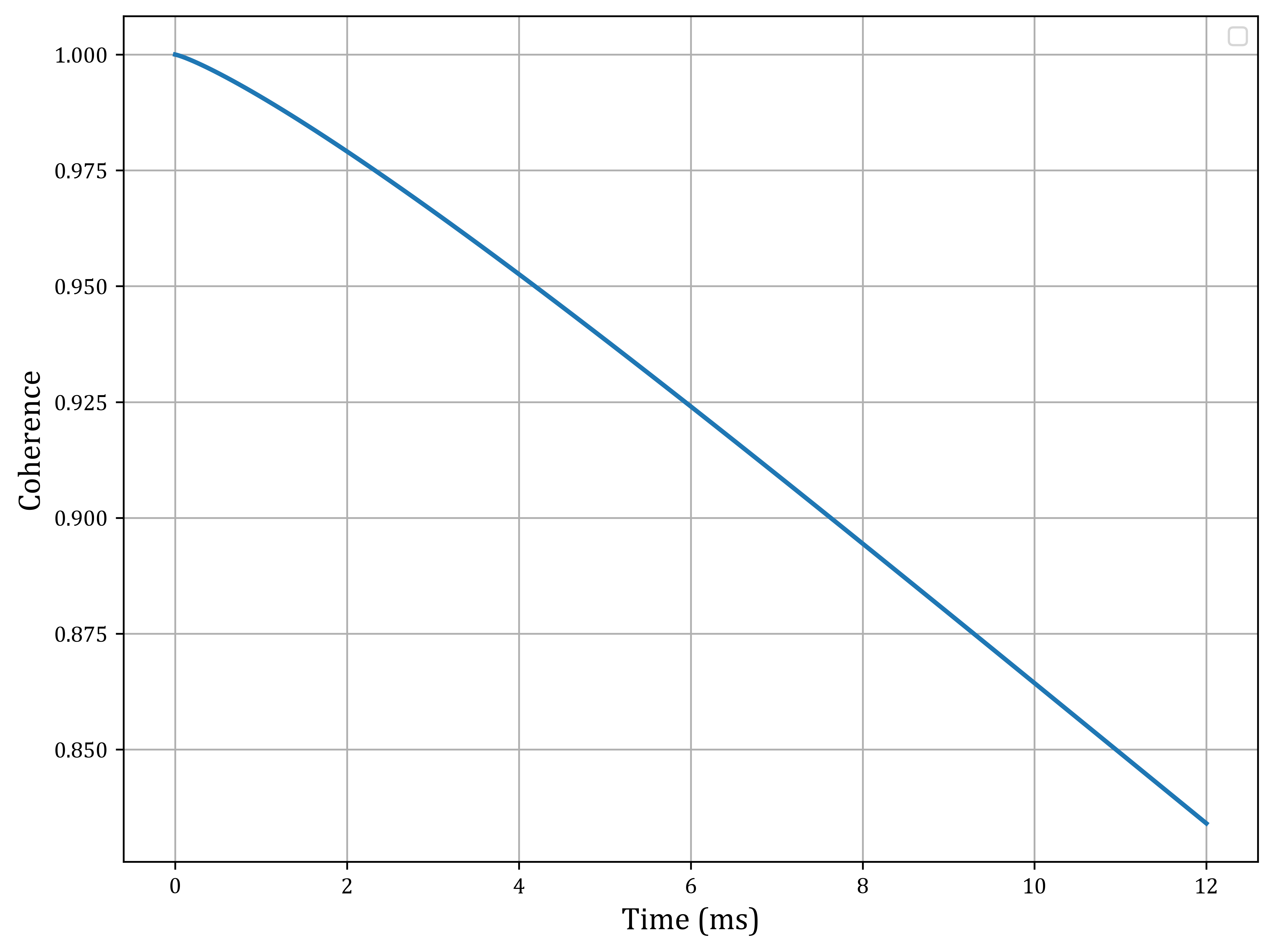}%
\label{fig:coherence}}

\caption{Noise characterization and phase stability of the trapped-ion qudit processor.
(a) Measured qudit-state population during repeated CHSH executions over a 12-ms interval. Blue circles show measured populations with $1\sigma$ error bars; the dashed line marks the mean, with dark/light shading indicating $\pm1\sigma$ and $\pm2\sigma$ bands. The orange curve is a 10-point moving average. Bounded fluctuations with no significant long-term drift indicate stable coherent operation, with residual variation dominated by photon-counting statistics and weak low-frequency phase noise.
(b) Overlapping Allan deviation of the extracted phase fluctuations (bootstrap uncertainties, 500 moving-block resamples; $\tau\lesssim N/5\times\Delta t$ for adequate window independence). The short-time regime ($\tau\lesssim0.6$~ms) shows a well-resolved decreasing trend consistent with white phase noise ($\mu_1\approx-0.4$, $r^2\approx0.80$). At longer $\tau$ the curve plateaus; a fit gives $\mu_2\approx-0.02$ with $r^2\approx0.05$, statistically indistinguishable from flat given the 120-point, 12-ms record, so this regime is read as inconclusive rather than confirmed drift.
(c) Phase-noise power spectral density via Welch's method. With only $\sim$2 independent segments at this record length, the fitted exponent is indicative rather than precise; a longer time series is needed to resolve the low-frequency slope reliably.
(d) Ramsey-type coherence decay, showing a stretched-exponential form characteristic of weak stochastic phase diffusion. Together, these measurements demonstrate high short-time phase stability, while long-time decoherence and its spectral character await longer data sets.}
\label{fig:noise_characterization}
\end{figure*}
\subsection{Qudit initialization}

The qudit is initialized by polarization-selective optical pumping into either $|q_0\rangle$ or $|q_3\rangle$, depending on the protocol. Initialization into $|q_0\rangle$ is used whenever possible because it provides the highest preparation fidelity.

The state-preparation fidelity, $F = 1 - P_{\mathrm{error}}(t)$, was characterized by measuring the residual population outside the target qudit state as a function of optical pumping duration (Fig.~\ref{fig:fidelityfit}). The optical pumping dynamics are well described by a single-exponential model,
\[
P_{\mathrm{error}}(t) = A e^{-t/\tau} + C,
\]
where $A$ is the initial unpumped population, $\tau$ is the characteristic optical pumping timescale, and $C$ is the asymptotic steady-state error. Analysis of the fit residuals (Fig.~\ref{fig:fidelityfit}, inset) indicates that remaining infidelity is dominated by systematic effects—such as residual beam polarization impurities and quantization axis fluctuations—rather than statistical uncertainty. At long pumping durations, the residual population is suppressed to the $10^{-3}$ level, achieving a state-preparation fidelity of $99.8\%$ and establishing robust qudit initialization.
\begin{table*}[t]
\centering
\caption{CHSH parameter $S$ as a function of $\theta$.}
\label{tab:chsh_theta}
\renewcommand{\arraystretch}{1.1} 
\setlength{\tabcolsep}{6pt}
\small

\begin{tabular}{c|c|c|c|c|c|c|c}
\hline\hline
$\theta$ (rad) & 0.111 & 0.331 & 0.661 & 0.991 & 1.32 & 1.65 & 1.98 \\
\hline
$S$ & $1.69\pm0.08$ & $1.42\pm0.08$ & $0.32\pm0.09$ & $-0.54\pm0.09$ & $-1.23\pm0.08$ & $-1.46\pm0.09$ & $-2.76\pm0.08$ \\
\hline
\end{tabular}

\vspace{0.15cm} 

\begin{tabular}{c|c|c|c|c|c|c|c}
\hline
$\theta$ (rad) & 2.31 & 2.64 & 2.97 & 3.30 & 3.63 & 3.96 & 4.30 \\
\hline
$S$ & $-2.82\pm0.08$ & $-2.62\pm0.07$ & $-1.88\pm0.09$ & $-1.34\pm0.09$ & $-0.54\pm0.09$ & $0.16\pm0.08$ & $0.90\pm0.08$ \\
\hline
\end{tabular}

\vspace{0.15cm} 

\begin{tabular}{c|c|c|c|c|c|c|c}
\hline
$\theta$ (rad) & 4.63 & 4.96 & 5.29 & 5.62 & 5.95 & 6.28 & 6.61 \\
\hline
$S$ & $1.40\pm0.07$ & $2.08\pm0.09$ & $2.41\pm0.09$ & $2.55\pm0.08$ & $2.13\pm0.09$ & $1.50\pm0.10$ & $0.84\pm0.08$ \\
\hline\hline
\end{tabular}
\end{table*}

\begin{table*}[t]
\centering
\caption{Measured-state probabilities (mean $\pm$ uncertainty) for Grover's algorithm.}
\label{tab:grover_confusion}
\small
\begin{tabular}{c|c|c|c|c}
\hline\hline
\textbf{Measured State} & \multicolumn{4}{c}{\textbf{Target State}} \\
\cline{2-5}
 & $|q_0\rangle$ & $|q_1\rangle$ & $|q_2\rangle$ & $|q_3\rangle$ \\
\hline
$|q_0\rangle$ & $0.942 \pm 0.016$ & $0.025 \pm 0.018$ & $0.019 \pm 0.010$ & $0.033 \pm 0.019$ \\
$|q_1\rangle$ & $0.028 \pm 0.020$ & $0.945 \pm 0.020$ & $0.026 \pm 0.015$ & $0.016 \pm 0.016$ \\
$|q_2\rangle$ & $0.048 \pm 0.015$ & $0.014 \pm 0.014$ & $0.848 \pm 0.033$ & $0.120 \pm 0.028$ \\
$|q_3\rangle$ & $0.061 \pm 0.021$ & $0.008 \pm 0.011$ & $0.107 \pm 0.030$ & $0.868 \pm 0.032$ \\
\hline\hline
\end{tabular}
\end{table*}

\subsection{Noise analysis and phase stability}

Repeated qudit-population measurements were recorded during the CHSH protocol over $8$--$10~\mathrm{ms}$ to characterize temporal stability. Figure~\ref{fig:noise_characterization}a shows the populations remaining bounded around $\sim0.312$ with no abrupt loss of contrast, indicating stable coherent operation.

Populations were converted to an accumulated phase via $P(t)=(1+\cos\phi(t))/2$, giving $\phi(t)=\arccos[2P(t)-1]$; the resulting phase trajectories show weak stochastic wandering, consistent with slow diffusion rather than rapid randomization.

Temporal stability was quantified via the overlapping Allan deviation~\cite{allan1966},
\begin{equation}
\sigma_A(\tau)=
\sqrt{\frac{1}{2}\left\langle\left(\bar{\phi}_{k+1}(\tau)-\bar{\phi}_{k}(\tau)\right)^2\right\rangle},
\label{eq:allan}
\end{equation}
with uncertainties from a moving-block bootstrap (500 resamples) to account for the overlapping estimator's reuse of samples. As shown in Fig.~\ref{fig:noise_characterization}b, the short-time regime ($\tau\lesssim0.6$~ms) follows $\sigma_A(\tau)\propto\tau^{\mu_1}$ with $\mu_1\approx-0.4$ ($r^2\approx0.80$) --- characteristic of white phase noise~\cite{allan1966,rubiola2008} from projection and photon-counting statistics, and the main robust result of this analysis. At longer times the curve plateaus rather than following a resolvable power law ($\mu_2\approx-0.02$, $r^2\approx0.05$); given the $N=120$-sample, 12-ms record, this is statistically indistinguishable from flat, and we do not interpret it as evidence of drift.

The phase-noise power spectral density~\cite{welch1967} (Fig.~\ref{fig:noise_characterization}c) shows enhanced low-frequency weight, qualitatively consistent with flicker-type ($1/f^\alpha$) noise~\cite{paladino2014}, but with only $\sim$2 independent segments at this record length the exponent $\alpha$ is not resolved with confidence.

The phase-diffusion coefficient $D_\phi=\mathrm{Var}[\phi_{i+1}-\phi_i]/2\Delta t$ gives an effective coherence time $T_2^{*}\approx1/(2D_\phi)$. The coherence envelope, modeled as a stretched exponential $C(t)=\exp[-(t/T_2^{*})^\beta]$ (Fig.~\ref{fig:noise_characterization}d), decays gradually, consistent with a mixed white/correlated-noise environment~\cite{wang2020qudit, ringbauer2022}.

In summary, the data firmly establish high short-time phase stability dominated by white phase noise; the PSD and coherence decay are qualitatively consistent with additional low-frequency noise, but neither the long-time Allan exponent nor the PSD slope is resolved with statistical confidence at the present record length, motivating longer or more densely sampled measurements.
\end{document}